\documentstyle[aps,pre,twocolumn,epsf]{revtex}
\def\B.#1{{\bbox{#1}}}
\begin{document}
\title{Dynamical equations for high-order structure functions, and
a comparison of a mean field theory with experiments in
three-dimensional turbulence}
\author {Susan Kurien and Katepalli R. Sreenivasan\\Physics Department and
Mason Laboratory, Yale University, New Haven, CT 06520-8286}
\maketitle
\begin{abstract} Two recent publications [V. Yakhot, Phys. Rev. E {\bf 63}, 026307,
(2001) and R.J. Hill, J. Fluid Mech. {\bf 434}, 379, (2001)]
derive, through two different approaches that have the
Navier-Stokes equations as the common starting point, a set of
steady-state dynamic equations for structure functions of
arbitrary order in hydrodynamic turbulence. These equations are
not closed. Yakhot proposed a ``mean field theory" to close the
equations for locally isotropic turbulence, and obtained scaling
exponents of structure functions and an expression for the tails
of the probability density function of transverse velocity
increments. At high Reynolds numbers, we present some relevant
experimental data on pressure and dissipation terms that are
needed to provide closure, as well as on aspects predicted by the
theory. Comparison between the theory and the data shows varying
levels of agreement, and reveals gaps inherent to the
implementation of the theory.
\end{abstract}

\section{Introduction}
It is well known that the Navier-Stokes (NS) equations of fluid
motion can be written in terms of statistical quantities such as
the moments of turbulent velocity at several simultaneous spatial
points. This statistical reformulation of the dynamic equations
introduces extra variables, giving rise to the familiar `closure'
problem in hydrodynamic turbulence. If the interest is in the
small-scale properties of turbulence, it is more appropriate to
obtain equations for the so-called structure functions, which are
the moments of velocity increments $\Delta \B.u_r =
\B.u(\B.x+\B.r) - \B.u(\B.x)$ over a separation vector, $\B.r$.
Such an equation for the third-order structure functions is known
from Kolmogorov's pioneering work \cite{kolm41a}. This equation is
unique because it is exact for homogeneous turbulence. For the
so-called longitudinal velocity increment $\Delta u_r = u(x+r) -
u(x)$, where $u$ is the velocity component along the direction of
the separation vector, the third-order structure function is given
by
\begin{equation}
\langle \Delta u_r^3 \rangle = -\frac{4}{5} \langle \varepsilon
\rangle r,
\end{equation}
where $\langle \varepsilon \rangle$ is the average energy
dissipation rate. For the third-order structure function of the
so-called transverse velocity increment $\Delta v_r = v(x+r) -
v(x)$, where $v$ is the velocity component transverse to the
separation vector, we also have the result
\begin{equation}
\langle \Delta v_r^3 \rangle = 0.
\end{equation}
The inertial range is defined by $\eta << r << L$, where $\eta$ is
the Kolmogorov scale and $L$ is the large scale of turbulence.
Equations (1) and (2) are both parts of Kolmogorov's 4/5-ths law.

Recently, Yakhot \cite{yakhot1} and Hill \cite{hill1} have derived
dynamical equations for high-order structure functions. Yakhot
first derived an equation for the so-called generating function
$Z$ from which structure functions of all orders can be obtained
by simple differentiation. Hill used a more conventional approach
to generate structure function equations. The equations are new,
and it is therefore useful to examine if anything further can be
learnt about turbulence through them. However, unlike Kolmogorov's
4/5-ths law, these equations are not closed. Yakhot presents a
``mean-field approach" to obtain pressure and dissipation
contributions, thereby closing the equations. From these closed
equations, it is possible to obtain certain small-scale properties
such as the probability density function (PDF) of transverse
velocity differences and the scaling exponents of structure
functions of all orders. Our goal here is to discuss these new
equations for high-order structure functions so as to clarify the
closure assumptions, and assess the mean field approach by
providing experimental comparisons for theoretical predictions.

Since the equations and the procedure for deriving them are not
yet familiar, we summarize them in Sec.~II and, for later use,
explicitly write them down for structure functions of several
orders. Section III introduces the experimental background needed
for our purposes, while Sec.~IV examines the approximate balance
of the equations without closure assumptions---mostly to set the
stage for further discussions. We summarize the mean field theory
in Sec.~V and present in Sec.~VI comparisons of its predictions
with experimental data on the PDFs of transverse velocity
increments and their scaling exponents. Section VII deals more
explicitly with the magnitude of dissipation terms, and our
conclusions are summarized in Sec.~VIII.

\section{Theoretical background}
\subsection{Brief review of the relevant equations}
Yakhot \cite{yakhot1} writes the NS equations in terms of the
generating function $Z = <e^{\lambda \cdot \Delta \B.u_r}>$,
where $\Delta \B.u_r$ is the vector velocity difference between
two space points $\B. x_1$ and $\B. x_2$ separated by the vector
distance $\B. r$. The generating function $Z$ is constructed, in
the spirit of field theories, so that its Laplace transform gives
the probability density function of the velocity differences, and
obeys the equation
\begin{equation}
{\partial Z \over \partial t} + {\partial^2 Z \over \partial
\lambda_\mu
\partial r_\mu} = I_f + I_p + D \label{z_eq}.
\end{equation}
Here $I_f$, $I_p$ and $D$ are the (known) forcing, pressure and
dissipative terms respectively, and are given by
\begin{eqnarray}
I_f & = & \langle \B.\lambda \cdot \Delta \B.f e^{\lambda \cdot
\Delta \B.u_r} \rangle \\ \nonumber I_p & = & -\B.\lambda \cdot
\langle e^{\lambda
\cdot \Delta \B.u_r}(\nabla_2 p(\B.x_2) - \nabla_1 p(\B.x_1)) \rangle \\
\nonumber D &=& \nu \B.\lambda \cdot \langle (\nabla_2^2 \Delta
\B.u_r(\B. x_2) -\nabla_1^2 \Delta \B.u_r (\B. x_1)) e^{\lambda
\cdot \Delta \B.u_r} \rangle.
\end{eqnarray}
The forcing term in the inertial range is small and may be
neglected. The closure of the equation requires a knowledge of
$I_p$ and $D$. In particular, the advection terms are treated
exactly. The pressure term $I_p$ contains correlation functions of
the form $<(\Delta \B.u_r)_i (\Delta \B.u_r)_j ...(\Delta
\B.u_r)_m \Delta(\nabla p)>$, and so one requires only the
knowledge of the correlation of pressure gradient $increments$ and
multipoint velocity increments. The multipoint energy dissipation
function $D$ has a structure that depends on details such as the
order of the moment considered. As we shall see later, its
structure resembles the well-known refined similarity hypotheses
of Kolmogorov \cite{kolm62}. In any case, the terms needed for the
closure of Eq.~(\ref{z_eq}) are, in principle, well-defined.

In homogeneous, isotropic turbulence, the following transformation
of variables is justified: $h_1 = r$, the separation in the
direction parallel to the separation vector; $h_2={\lambda.\B.{r}
\over r}$, the component of $\lambda$ along $r$; $h_3 =
\sqrt{\lambda^2 - h_2^2}$, the component of $\lambda$ in the
direction perpendicular the separation vector. In these new
variables, Eq.~(\ref{z_eq}) for the generating function becomes
\begin{eqnarray}
\partial_t Z + [\partial_1 \partial_2 + {d - 1 \over r} \partial_2 +
{h_3 \over r} \partial_2 \partial_3 &+& \\\nonumber {(2 -
d)h_2 \over rh_3}\partial_3 - {h_2 \over r}\partial_3^2]Z
&=& I_f + I_p + D \label{z_eq2},
\end{eqnarray}
where $\partial_i$ denotes the partial derivative with respect to
$h_i$. In the new variables, the generating function can be
written as
\begin{equation}
Z = <e^{h_2 \Delta u_r + h_3 \Delta v_r}>
\end{equation}
where $\Delta u_r$ and $\Delta v_r$ are the familiar velocity
differences in the longitudinal and transverse directions,
respectively. The structure functions are then generated by
successive differentiation of $Z$ as
\begin{equation}
S_{n,m} \equiv \langle (\Delta u_r)^n (\Delta v_r)^n \rangle =
\partial_2^n \partial_3^m Z(r)\vert_{h_2=h_3=0}.
\end{equation}
Let us first focus attention on even-order structure functions for
which the dissipation terms are negligible in the inertial range.
Multiply Eq.~(\ref{z_eq2}) by $h_3$, perform $\partial_3
\partial_2^{2n-1}$ of the resulting equation, and take the limit
$h_2 = h_3 \to 0$. We then have
\begin{eqnarray}
{\partial S_{2n,0} \over \partial r} &+& {d-1 \over r}S_{2n,0} -
{(d-1)(2n-1) \over r}S_{2n-2,2} = \\\nonumber &-&(2n-1)\langle
\Delta p_x(\Delta u)^{2n-2}\rangle  \\\nonumber &+& P(1 -
\cos{r/l_f})a_{n,d}S_{2n-3,0}, \label{S_2n0}
\end{eqnarray}
where $d$ denotes the space dimension, $P$ is the mean energy
``pumping" rate, $l_f$ is the forcing scale and $a_{n,d} \equiv
{2(2n-1)(2n-2) \over d}$. For $n=1$, we obtain the well-known
relationship between the second-order longitudinal and transverse
structure functions, as
\begin{equation}
{\partial S_{2,0} \over \partial r} + {d - 1 \over r}S_{2,0} = {d
- 1 \over r}S_{0,2}.
\end{equation}
For $n=2$, Eq.~(\ref{S_2n0}) yields a new relation
\begin{equation}
{\partial S_{4,0} \over \partial r} + {d-1 \over r}S_{4,0} =
{3(d-1) \over r}S_{2,2} - 3 \langle (\Delta p_x)(\Delta u)^2
\rangle, \label{S_40_ex}
\end{equation}
which is exact in the case of incompressible, isotropic turbulence
in the limit of zero viscosity. To extract further information
from this equation, one has to invoke some closure. From now on,
we consider only three dimensional turbulence unless specified
otherwise. If the pressure term is small in the inertial range (a
consideration to which we will return momentarily), we obtain
\begin{equation}
{\partial S_{4,0} \over \partial r} + {2 \over r}S_{4,0} \approx
{6 \over r}S_{2,2}. \label{S_40_approx}
\end{equation}

Following this same procedure, it is now easy to write down a
sixth order equation with $n=3$ in Eq.~(\ref{S_2n0}). Neglecting
the pressure term again, we obtain
\begin{equation}
{\partial S_{6,0} \over \partial r} + {2 \over r}S_{6,0} \approx
{10 \over r}S_{4,2}. \label{S_60_approx}
\end{equation}
In a similar manner, we can extract two additional relations for
fourth and sixth orders. Their corresponding approximate forms
(again neglecting pressure contributions) are
\begin{equation}
{\partial S_{2,2} \over \partial r} + {4 \over r}S_{2,2} \approx
{4\over3r}S_{0,4} \label{S_04_approx}
\end{equation}
\begin{equation}
{\partial S_{2,4} \over \partial r} + {6 \over r}S_{2,4} \approx
{6\over5r}S_{0,6}. \label{S_06_approx}
\end{equation}

Equations (\ref{S_40_approx})-(\ref{S_06_approx}) were also
derived by Hill \cite{hill1} by making a convenient change of
variables so that the scale-dependent parts of the equation can be
separated from the position-dependent parts and by eliminating the
position-dependence in the presence of homogeneity. For the
particular case of isotropy, Hill developed a matrix algorithm for
solving the system of equations that determined all components of
the tensorial structure function of a given order. The dynamical
equations obtained for homogeneous and isotropic turbulence
confirm Yakhot's results. Hill's formulation provides an
additional useful fact that there are exactly two equations
relating fourth-order structure functions, namely
Eqs.~(\ref{S_40_approx}) and (\ref{S_04_approx}). For the sixth
order, his procedure shows that there are $three$ equations
relating the non-zero components of structure functions, and the
third equation is easily generated. Again without the pressure
terms, this remaining sixth-order equation is
\begin{equation}
{\partial S_{4,2} \over \partial r} + {4 \over r}S_{4,2} \approx
{4\over r}S_{2,4}. \label{S_mixed6_approx}
\end{equation}
Equations (\ref{S_40_approx})-(\ref{S_mixed6_approx}) are
relationships among different components of structure function
tensor of the same order. Without pressure terms, they form a
closed set for each even order.

Hill's procedure is convenient for writing down the equations for
odd-order structure functions. The equation for the third-order is
the well-known Kolmogorov equation (see Eqs.~(1) and (2)), which
need not be written down again. There are three equations for the
fifth order, which, again without pressure and dissipation terms,
are
\begin{equation}
{\partial S_{5,0} \over \partial r} + {2 \over r}S_{5,0} \approx
{8\over r}S_{3,2}; \label{S_50_approx}
\end{equation}
\begin{equation}
{\partial S_{3,2} \over \partial r} + {4 \over r}S_{3,2} \approx
{8\over 3r}S_{1,4}; \label{S_32_approx}
\end{equation}
\begin{equation}
{\partial S_{1,4} \over \partial r} + {6\over r}S_{1,4}\approx 0.
\label{S_14_approx}
\end{equation}

The $\approx$ symbol in these equations has to be treated with
greater caution than for the even-order case because the equations
for odd-order structure functions contain dissipation terms as
well, see \cite{yakhot1}. Therefore the approximation implied in
the equations (\ref{S_50_approx})-(\ref{S_32_approx}) implies the
neglect of dissipation terms as well. We will assess this
additional aspect in Sec.~VII, but can see by inspection that the
approximations cannot be correct for at least the last equation of
the above set. It contains only one component of the fifth order
structure function, $S_{1,4}$, which, when estimated using
Kolmogorov's K41 scaling \cite{kolm41b}, shows that the the
approximation Eq.(\ref{S_14_approx}) cannot hold true.

A further discussion of these equations, and the of degree to
which they may be reasonable, requires some contact with
experiments. It is therefore necessary to introduce some basic
experimental details at this stage before resuming the discussion
of the theory.

\section{Experimental conditions}
The velocity data were acquired by means of a $\times$-wire probe
mounted at a height of about 35 m above the ground on a
meteorological tower at the Brookhaven National Laboratory. The
hot wires were about 0.7 mm in length at 5 $\mu$m in diameter.
They were calibrated just prior to being mounted on the tower, and
operated on DISA 55M01 constant-temperature anemometers. The
frequency response of the hot wires was typically good up to 20
kHz. The voltages from the anemometers were low-pass filtered and
digitized. The low-pass cutoff was never more than half the
sampling frequency $f_s$. The sampling rate was adequate to
resolve most of the scales, including dissipative ones. The
voltages were converted to velocities in a standard way through
the standard nonlinear calibration procedure. The mean wind
velocities, roughly constant over the duration of a given data
set, ranged between 5 and 10 ms$^{-1}$ in the entire series. The
usual procedure of surrogating time for space (``Taylor's
hypothesis'') was used to obtain the mean dissipation rate
$\langle \varepsilon \rangle$ and an estimate for the Kolmogorov
scale $\eta$.

The relevant details of the data analyzed here are given in
\cite{bd1,bd2}. Briefly, the Taylor microscale Reynolds number is
10,680, the large scale $L$ is about 42 m, Kolmogorov scale $\eta$
is 0.44 mm, the scaling range according to the linear part in the
third-order structure function is conservatively between 0.01 m
and 0.2 m (although it could be stretched in either direction by
factors of 2). This will be regarded as the operational definition
of the inertial range.

The theoretical development discussed here is meant for locally
isotropic turbulence appropriate to asymptotically large Reynolds
numbers. The present measurements are indeed at large enough
Reynolds numbers, but it is not obvious that the small scales are
isotropic. Indeed, we have used similar data before
\cite{arad,kurien} to extract anisotropic parts of the structure
functions by performing the SO(3) decomposition. Thus, a few
explanatory words regarding the degree of isotropy are appropriate
here.

Figure \ref{bd_spect} shows the longitudinal spectral density
$E_{11}$ and the transverse spectral density $E_{22}$ for one of
the data sets. If strict isotropy prevails in the inertial range,
the ratio $E_{22}/E_{11}$ should be 4/3. The data show that the
ratio, while varying slowly in the inertial range from a somewhat
smaller value than 4/3 to a somewhat larger value, is not far from
being 4/3. Similarly, for third-order structure functions, we have
the exact result \cite{monin} that the ratio $S_{1,2}/S_{3,0}$
should be 1/3. Measurements show that this ratio, which is indeed
reasonably constant in the inertial range, has a magnitude of
about 0.42 (see Fig.~{\ref{s12bys3}). There is undoubtedly some
degree of anisotropy in the inertial range, and so the conclusions
are to some degree affected by this artifact. To pursue this issue
further, we note that $S_{3,0}$ has the expected value of 4/5, and
$S_{0,3}$, which should be zero exactly for the isotropic case, is
of the order of 0.1 $S_{3,0}$. The anisotropy is thus not large,
and the transverse component is perhaps the more anisotropic. This
conclusion is consistent with \cite{kurien,biferale}.

In summary, then, we regard for present purposes the departures
from isotropy to be relatively small in the inertial range,
especially for even orders, and expect that the results will not
be qualitatively affected by their presence. The quantitative
effect is not easy to ascertain, though it is likely to be modest
from the considerations just cited.

\section{The balance of the approximate equations}
Using the experimental data just described, we calculate the
left-hand side (LHS) and right-hand side (RHS) of each approximate
equation of Sec.~II, and obtain the relative size of the
difference (LHS - RHS)/LHS; for even orders, this difference is an
estimate of the magnitude of the neglected pressure terms. Figures
\ref{s40}, \ref{s60}, \ref{s04}, \ref{s06} and \ref{sm6} display
the results for Eqs.~(\ref{S_40_approx}), (\ref{S_60_approx}),
(\ref{S_04_approx}), (\ref{S_06_approx}) and
(\ref{S_mixed6_approx}), respectively. The absolute value of the
relative difference is also shown in each case. In the inertial
range, all four equations seem to balance reasonably well without
the pressure terms, their contribution being of the order of
10$\%$ of the LHS for all equations, growing larger as $r$ becomes
larger as well as smaller. To that extent, the closure of
Eqs.~(\ref{S_40_approx})-(\ref{S_mixed6_approx}) in the inertial
range seems to be approximately justified.

The situation with respect to odd-order structure functions is
somewhat different. Odd-order moments do not converge as well as
even-order moments do, and so we may expect significantly more
scatter in experimental plots. The terms for the three equations
(\ref{S_50_approx})-(\ref{S_14_approx}) are shown in
Figs.~\ref{s50}-\ref{s14}. Despite the large scatter, it is
reasonably clear that the relative difference of LHS and RHS in
Fig.~\ref{s50} is of the order of $20-30\%$ in the inertial range.
In Fig.~\ref{s32}, the relative difference is perhaps larger, up
to $50\%$ at places. For Eq.~(\ref{S_14_approx}), as was expected
and remarked upon earlier, there is no balance at all (see
Fig.~\ref{s14}).

Before examining these issues, it is worth commenting on the
intriguing observation that the pressure effects are small in the
inertial range {\it at least for even-order structure functions}.
This feature suggests that the operating physics there might have
some relation to that of the forced Burgers equation. It further
suggests that the pressure terms may become effective only in the
dissipation range (and also for large scales, which we will not
consider here). It is reasonable to suppose that vortex structures
(qualitatively like wing-tip vortices) are generated as soon as
the pressure terms are activated. Perhaps the small-scale vortex
tubes (also called worms, see \cite {mene,jime}), are a result of
this effect. Clearly, the validity of the pressureless physics is
limited because the scaling exponents saturate for the forced
Burgers equation, whereas there is no evidence that the saturation
occurs for three-dimensional turbulence (at least for moments up
to order ten; see Sec.~VI). In any case, the smallness of pressure
terms suggests that the intercomponent energy transfer (for which
they are responsible) is small, and that any remnant anisotropies
at the large-scale end of the inertial range tend to persist, or
at least not diminish rapidly, as the scale size decreases through
the inertial range. There is growing evidence from other
measurements that this might indeed be so \cite{kurien,warhaft}.

\section{A mean field theory}
The previous section has shown that the imbalance in the
approximate equations for even-order structure functions, caused
entirely by the neglect of pressure terms, is of the order 10\%.
The imbalance is larger for odd-orders, for which it is to be
remembered that dissipation terms are also important. Thus,
ignoring pressure and dissipation terms is not an option in
general, and it is clear that one must make a plausible theory for
them. This has been attempted in the mean field theory of
\cite{yakhot1}. Though our interest and contributions are
primarily in the experimental assessment of the theory, it is
helpful to provide here a summary of the theory itself---if only
to clarify the motivation for the experimental tests performed. We
shall focus on the essence of the physical arguments, rather than
on analytical details.

\subsection{General remarks}
Mean field theories provide approximate means of describing a
thermodynamic system by supposing that each `particle' in a
many-body system moves in the `mean' field of all other particles
in the system. This is opposite to the situation in which only
nearest neighbor interactions matter. More formally, attribute to
the system an order parameter $\phi$ that is zero when the system
is ordered and becomes increasingly non-zero with increasing
disorder. If the fluctuations in the order parameter are small,
then it may be replaced by a spatially uniform average value. The
mean field approximation implies infinite range interactions;
while this cannot be realized in practice, the order parameter in
many thermodynamic systems could become arbitrarily small as the
temperature approaches a phase transition value, $T_c$. The
Ginzburg-Landau theory makes use of this feature to propose a
description of the free energy and to derive critical exponents at
phase transitions. In general terms, the free energy $F(\phi, T)$
is expanded in powers of $\phi$ as
\begin{equation}
F = F_0 + A\phi + B\phi^2 + C\phi^3 + \dots
\end{equation}
where $A$, $B$ and $C$ are functions of $T$. Near the critical
point in the $T$-space, where $T \rightarrow T_c$, the expansion
can be truncated at the lowest order terms in $\phi$. The
expansion then provides a qualitative description of the
thermodynamic processes; in practice, this mean field approach may
work even far from the critical point.

Strictly speaking, a mean field theory may not apply to turbulence
where quantities such as the free energy and order parameter
cannot be defined unambiguously. In Yakhot's theory, the idea is
carried over qualitatively by identifying a small parameter in
some regime and expanding other dependent quantities around that
small parameter. The `phase transition' considered is the change
of sign in energy flux that occurs in going from two-dimensional
($2d$) to three-dimensional ($3d$) turbulence. It is understood
from Kolmogorov's equation for the third-order structure function
that the energy transfer is from the small to the large scale in
$2d$ turbulence, and vice versa in $3d$ turbulence. It is assumed
that this change is continuous, changing sign at some critical
dimension $d_c$---analogous to the critical temperature $T_c$ in
thermodynamic phase transitions. In $2d$ turbulence, the
dissipation is negligible for high Reynolds numbers (because the
energy ultimately concentrates in the large scale). In $3d$
turbulence, on the other hand, the dissipation is the key to
energy transfer from large to small scales. The hope, then, is
that both pressure and dissipation can be expanded in terms of a
small parameter in the vicinity of $d_c$.

\subsection{The pressure terms}
In turbulence dynamics, transverse structure functions do not
participate in energy transfer. Yakhot therefore regards the
fluctuations in the transverse velocity increment $\Delta v_r$ as
small---in effect, if not in actual fact. It is known from
numerical simulations \cite{boffetta} as well as experiments
\cite{tabeling} of the inverse cascade in $2d$ turbulence that
$\Delta v_r$ is almost exactly gaussian. The absence of
intermittency makes it plausible to regard the fluctuations as
``small". We shall therefore consider the $2d$ case briefly.

The key step for further analysis is the introduction of a
conditional expectation of the pressure gradient increment for a
fixed value of $\Delta u_r$, $\Delta v_r$ and $r$ as
\begin{eqnarray}
\langle \partial_y p(x+r) &-& \partial_y p(x) | \Delta u_r, \Delta
v_r, r\rangle \\\nonumber &\approx& \sum_{m,n}
\kappa_{m,n}(r)(\Delta u_r)^m(\Delta v_r)^n.
\end{eqnarray}
This is related to the needed correlations in $I_p$ which is of
the form
\begin{eqnarray}
&\langle& (\partial_y p(x+r) - \partial_y p(x)) (\Delta u_r)^p
(\Delta v_r)^q \rangle \\ \nonumber &=& \int \langle \partial_y
p(x+r) -
\partial_y p(x) | \Delta u_r, \Delta v_r, r\rangle
\Delta u_r^p \Delta v_r^q \\\nonumber &\times& P(\Delta u_r,
\Delta v_r, r) d(\Delta u_r) d(\Delta v_r) d^3 r.
\end{eqnarray}
The use of the conditional expectation provides a tool for
expanding the pressure terms in terms of the ``small quantity"
$\Delta v_r$. Now, in the spirit of the Ginzburg-Landau expansion,
only the lowest order terms in $\Delta v_r$ are retained
(corresponding to $\Delta u_r \Delta v_r$ and $\Delta v_r$). The
prefactors of the expansion are constrained by the
incompressibility condition and by the dimensionality of space.

By substituting in Eqs.~(9)-(11) the pressure term derived from
the conditional expectation value, and assuming the exponents to
be given by from Kolmogorov's K41 scaling arguments
\cite{kolm41b}, Yakhot concludes that the high-order even moments
are consistent with gaussianity. The argument is circular but
internally consistent. The gaussianity of the transverse increment
$\Delta v_r$ is then deduced from Eqs.~(13)-(15). This is in
excellent agreement with the results of numerical simulations of
\cite{boffetta}. Thus, we might conclude that a plausible mean
field expression for the pressure contribution exists for $2d$.

The next crucial assumption of the theory is that the above form
of the mean field approximation is applicable also for $3d$
turbulence. The rationale is not easy to articulate, especially
because, unlike in $2d$ turbulence, the PDFs of $\Delta v_r$
possess stretched-exponential tails in $3d$ turbulence
\cite{sreeni}. We shall provide some statements of mild
justification subsequently, but emphasize that the validity of
this assumption has to rest on the basis of the agreement, or lack
thereof, with experiments.

\subsection{The small parameter and the dissipation term}
We need to consider the dissipation term before turning to
experiments again. In the inverse cascade range in $2d$ turbulence
the dissipation term $D$ can be set to zero because the flow
evolution is towards larger and larger scales. However, $D$ is
central in $3d$ turbulence, and it is known that dissipation
fluctuations are immense at high Reynolds numbers \cite{meneveau}.
The objective in a mean field approach is to locally smooth out
the fluctuations, through some procedure such as Obukhov's
\cite{obukhov}. For closure, there is a need to relate this
coarse-grained dissipation field to velocity fluctuations,
analogous to that employed in Kolomogorov's refined similarity
hypotheses \cite{kolm62,stolo}. Yakhot's theory is similar in
spirit but the details are different, as we shall illustrate.

Let us denote a coarse-grained velocity field for a given spatial
scale $r$ by $V_r$. This will be assumed to be the same as $\Delta
v_r$. Certain one-loop calculations of Yakhot and Orszag
\cite{yakhot2} give the effective viscosity as
\begin{eqnarray}
\nu_r &\approx& (d-d_c)^{1 \over 3}N(\varepsilon_r r^4)^{1 \over
3}
\\\nonumber &\approx& V_r^2 \tau_r + higher~order~nonlinear~terms
\end{eqnarray}
where $N$ is a constant that depends weakly on the space
dimensionality, $d$, and $\varepsilon_r$ is the dissipation rate
coarse-grained on the scale $r$. If we ignore nonlinear terms,
this equation provides a natural definition of $\tau_r$, the
characteristic time for the field $V_r$.

There is no obvious justification for ignoring the higher order
nonlinear terms which, in $3d$ turbulence, are typically $O(1)$,
nor in assuming that $\tau_r$ is small compared to $V_r/r$.
However, if we assume that the theory can be analytically
continued into non-integer dimensions between 2 and 3, a suitable
small parameter can be generated as follows. The time scale
characterizing the interaction of a scale $r$ with all other
scales less than $r$ is the so-called eddy turn-over time, or the
time taken for energy transfer to occur between $r$ and the
Kolmogorov scale $\eta$. One may use K41 to estimate this time
scale. The process of energy transfer can be thought to consist of
two distinct steps, one involving nonlinear transfer across scales
without any pressure effects, and another involving the relaxation
due to pressure effects. In $3d$ turbulence, these two steps are
parts of the same simultaneous process, so the time scales
associated with them cannot be separated. But, if, as one
approaches $d_c$, it is increasingly true that the pressure
effects are small except when scales of the order $\eta$ are
reached, the two time scales involved could become disparate, and
the relaxation due to pressure terms enters the picture only at
the smallest scale and can therefore be assumed to be fast. Then
the dimensionless ratio $ q_r \equiv {\tau_r/\theta_r}$, where
$\tau_r$ is the time scale for relaxation effects and $\theta_r$
is the time scale for energy transfer, would be a small parameter.

Using this basic idea and his one-loop calculations
\cite{yakhot2}, Yakhot deduces the following results: $$ d_c =
2.56,~q_r \approx (d - d_c)^{1/2},~V_r \approx (d - d_c)^{-1/6},$$
and so forth. The notion of a critical dimension is not new (see
\cite{frisch}), though the estimates for it in \cite{yakhot1} and
\cite{frisch} are substantially different. The precise numbers and
powers in the above equation depend in detail on the approximation
made to compute them, and are presumably not final; they cannot,
in any case, be verified experimentally near the critical
dimension. Here, we merely wish to draw upon the general idea of a
critical dimension near which a small parameter can be defined,
and in whose vicinity the energy piles up (as shown by the last of
the three relations above: the energy is being pumped at a
constant rate but is being transferred neither upscale as in $2d$
nor downscale as in $3d$). These ideas allow Yakhot to truncate
the effective viscosity and write the dissipation $\varepsilon_r$
in terms of the lowest order terms in terms of the coarse-grained
velocity field {\it in the vicinity of $d_c$}:
\begin{equation}
\varepsilon_r \approx - {1\over2} {\partial \over
\partial r_i}[V_{ri}V^2_{rj} (1+ O(d-d_c))]
\end{equation}

Perhaps two additional remarks might be usefully made. First, the
coarse-grained velocity fluctuations become very large as the
critical dimension is approached, yet it may seem that the mean
field approximation proposed for pressure terms assumes that
fluctuations are small. To avoid confusion, it is important to
keep in mind the distinction between fluctuations in longitudinal
and transverse velocity increments. The velocity scale that blows
up is related to energy transfer, and hence the longitudinal
velocity component, but the component whose fluctuations are
supposed to remain small is the transverse velocity. The sense in
which those fluctuations are small is unclear (because they too
are intermittent in $3d$, see \cite{sreeni}), but the fact remains
that it takes no part in energy transfer and so its $effects$ are
thought to be ``small" in some rough sense. Since the pressure
effects are small, the intercomponent energy transfer is
inhibited, and so, once fluctuations in $\Delta v_r$ are small at
some scale, they will presumably remain small at others as well.
Secondly, in order to be able to truncate the energy dissipation,
the higher order viscosity terms have to decay faster than the
rate of blow-up of velocity fluctuations. This is indeed the case
above.

Now, keeping in the mind the symmetries of the NS equations, the
simplest form for the $V_r$ contributions to the dissipation rate
is
\begin{equation}
\langle \varepsilon \rangle \approx c(d) \Delta u_r \Delta v_r
{\partial \Delta v_r \over
\partial r}
\end{equation}
The coefficient $c(d)$ must reflect the change in going from $2d$
to $3d$ (zero dissipation to finite dissipation). This may not be
a smooth change (as in second-order phase transitions) because
$c(d_c)$ could well be singular (as in first-order phase
transitions). Yakhot assumes, however, that it is $O(1)$ for
$d-d_c > 0$. Then $D$ takes on a form similar to Kolmogorov's
refined similarity hypothesis, relating $\langle \varepsilon_r
\rangle$ with the third-order longitudinal structure function:
\begin{eqnarray}
D &\approx& c(d)h_3 \partial_{h_2}\partial_{h_3} \partial_r Z
\\\nonumber &\approx& c(d)h_3^3 \langle \Delta u_r \Delta v_r
{\partial \Delta v_r \over
\partial r} e^{h_2 \Delta u_r + h_3 \Delta v_r} \rangle \\\nonumber
&+& terms~neglected.
\end{eqnarray}
This enables the closure to be complete.

A non-trivial difficulty is the testing of the theory in
non-integer dimensions near $d_c$. At present, the consequences of
the theory can only be tested in $2d$ or $3d$. The extrapolation
to non-integer dimensions is not an intrinsic limitation of the
theory, but reflects the lack of experimental ingenuity at
present. It must, however, be noted that in shell models where an
interaction parameter can be tuned to change the direction of
energy transfer, one can make more reasonable contact with the
theory. Such comparisons have been attempted recently
\cite{jensen} and the results are encouraging. We have already
noted that the conclusions of the theory are consistent with
experiments and simulations of $2d$ turbulence. We shall examine
in the rest of the paper the extent to which the predictions of
the theory are applicable also to $3d$.

\section{Comparison of the theory with measurements in three-dimensional
turbulence}
\subsection{Probability density function of transverse velocity
increments} When the forms of $I_p$ and $D$ from previous sections
are substituted into the full structure function equations, one
can generate the following equation for the PDF of transverse
velocity increments $P(\Delta v_r,r) \equiv P(V,r)$:
\begin{equation}
{\partial P \over \partial r} + {1+3\beta \over 3r}{\partial \over
\partial V} V P - \beta {\partial \over \partial V}V
{\partial P\over\partial r} = 0. \label{vpdf_r_eq}
\end{equation}
Here $\beta \propto c(d)$. This equation is linear and can be
solved in principle, but the solution has no simple analytic form.
(For some discussion of this aspect, see \cite{kurien_th}.) For
small $V$, however, the equation admits a solution of the type
$r^{\kappa}F(V/r^{\kappa})$ with
\begin{equation}
P(V = 0,r) \propto r^{-\kappa},
\end{equation}
where, from Yakhot's theory,
\begin{equation} \kappa \equiv {1+3\beta\over3(1-\beta)}
\approx 0.4
\end{equation}
and $\beta \approx 0.05$. Using the fact that the variance
$\sigma_V$ of $V$ is expected to vary as $r^{0.35}$ (see
\cite{bd1}), we have the result
\begin{equation}P({V \over\sigma_V} \rightarrow 0,r) \equiv
\sigma_V P(V\rightarrow 0,r) \approx r^{-0.05}. \label{pdf_peak}
\end{equation}
We shall now test these predictions.

The precise measurement of the peak value of the PDFs from the
data must be done carefully because it is sensitive to the bin
width chosen around $V = 0$. In our measurements, the bin width
around $V = 0$ was gradually refined until the PDF value at the
origin no longer depended on the bin size. Figure \ref{find_peak}
shows that $P(V \rightarrow 0)$ at $r \approx 100\eta$ asymptotes
to a value of 0.64. The sharp ascent of the numbers for very small
values of the bin width is an artifact of the extreme narrowness
of the bin width, which results in false values when normalizing.
The procedure was repeated for several values of $r$. Figure
\ref{peakpdfs} shows the properly normalized PDF values for $V=0$
for different scales $r$ ranging from the Kolmogorov scale $\eta$
to the large scale $L$. The scaling exponent for this quantity is
$\approx -0.065$ in the inertial range, numerically about 25$\%$
larger than the theoretical value of -0.05. With this
experimentally derived scaling exponent, we can evaluate that
$\beta \approx 0.058$ compared to the theoretical value of $0.05$,
a 16$\%$ difference. Figure \ref{nearpeak} also shows that the
form $r^{\kappa}P(V/r^{\kappa})$ is essentially constant for small
$V$.

One can obtain the form of $F$ for large $V$ by a steepest descent
approximation (see \cite{kurien_th} for more details). The result
from \cite{yakhot1} is
\begin{equation}
F\propto {1\over\sqrt{\Omega(r)}}\exp\Big(-{(\ln\xi)^2 \over
\Omega(r)} \Big), \label{pdf_anl1}
\end{equation}
where $\xi = V L^{\beta\kappa\over\beta - 1}/
r^{\kappa(1-2\beta)/1-\beta}$, and $\Omega(r) =
{\beta\kappa\over(1-\beta)^2}|\ln(r/L)|$. (The corresponding
expressions in \cite{yakhot1} are printed incorrectly.) The
prefactor of Eq.~(\ref{pdf_anl1}) is possibly $r$-dependent.
Equation (\ref{pdf_anl1}) can be re-written as
\begin{equation}
(-\Omega(r) \ln(P(V,r) r^\kappa \sqrt{\Omega(r)})^{0.5} \propto
\ln(\xi). \label{pdf_anl2}
\end{equation}
Figure \ref{pdf_ncoll} shows plausible linear behaviors for the
tails of the PDF in the proposed logarithmic units of
Eq.~(\ref{pdf_anl2}). There is, however, evidently still some
$r$-dependence that precludes their collapse. We recall that
corrections to steepest descent approximations are often
logarithmic, but are difficult to calculate here analytically. We
assume a dependence of the form $(\ln(r/L))^\gamma$ for the
proportionality factor. Figure \ref{pdf_coll} shows a replot of
the data with the additional factor of $(\ln(r/L))^2$ multiplying
the PDF. The exponent $2$ was chosen because it collapses the data
best in the inertial range. (The one separation distance that does
not collapse belongs to the dissipation range.)

Our main conclusion so far is that the mean field models for
pressure and dissipation terms provide a way for closing the PDF
equation, and for solving it for the limiting situations. The
prediction is that, to first approximation, the tails of the PDF
of $\Delta v_r$ are lognormal. The experimental data suggest that
this might be so, but that an $r$-dependent contribution is
missing. It is at present not clear whether this missing aspect is
merely a correction to asymptotics, or corresponds to additional
terms in the mean field expansion, or is even more fundamental.

\subsection{The scaling exponents and the prospect of their
saturation} Seeking the solution to Eq.~(\ref{vpdf_r_eq}) under
the K41 constraint for the third-order structure functions and
assuming $S_{0,n} \propto r^{\zeta_n}$, Yakhot obtained the
following formula for the structure function exponents:
\begin{equation}
\zeta_n = {n(1+3\beta) \over 3(1+\beta n)}. \label{y_exp}
\end{equation}

Table I and Fig.~\ref{exps} show the calculated exponents and
compare them with those obtained from the Direct Numerical
Simulations, or DNS, data \cite{chen} as well as experiments. The
agreement is good for all orders, perhaps slightly better for the
DNS data for high order exponents.

Using probability density functions to define the statistical
quantities, we have (putting $U = \Delta u_r$) the conditional
expectation value of $V^2$ for a fixed value of $U$, $Q_2(U)$, as
\begin{equation}
S_{2n-2,2} = \int P(U)U^{2n-2}Q_2(U)dU.
\end{equation}
See Sec.~IV of \cite{yakhot1}. The Kolmogorov scaling will hold
(by dimensional arguments) for $Q_2(U) \propto U^2$. On the other
hand, saturation of exponents, $\zeta_{2n} \rightarrow$ constant
as $n\rightarrow \infty$, is possible for $Q_2(U) \propto
U^{\delta}$ for $\delta < 2$ and $U$ large. We present the
conditional statistical quantity $Q_2(U)$ as a function of $r$ in
Fig.~\ref{q2plot}. It is not clear if the trend for large $U$ is
in agreement with the saturation condition. There is a very small
range of $U$ towards the tails which seems to vary as
${U^2/\ln(U^2)}$ but this is not conclusive. There might also be
the influence of anisotropy in the PDFs, as is evident, for
example, in the asymmetry of the joint PDFs, which in turn could
change the nature of the tails of the conditional statistics.

\section{Remarks on the magnitude of dissipation terms}
It is helpful to recall that the full form of the equation
relating even order transverse moments to mixed moments of the
same order in $3d$ is
\begin{eqnarray}
{\partial S_{2,2n} \over \partial r} + {2+2n \over r}S_{2,2n} &=&
{2+2n \over 2n+1}{S_{0,2n+2} \over r} \nonumber\\
&-& 2n\langle {\cal P}_{yv}\Delta u (\Delta v)^{2n-1}\rangle,
\label{S_02n+2_exact}
\end{eqnarray}
where ${\cal P}_{yv} \equiv \partial_y p(x+r) - \partial_y p(x)$.
The subscript $y$ denotes the component of the pressure gradient
while the subscript $v$ reminds us that this form of the pressure
may be expanded in $\Delta v_r$. In the inertial range, the
dissipation term is small (because the order of the structure
functions is even) and the forcing term negligible.

Compare this equation to the one relating the odd order mixed
moments
\begin{equation}
{\partial S_{1,2n} \over \partial r} + {2+2n \over r}S_{1,2n} = -
2n\langle {\cal P}_{yv}(\Delta v)^{2n-1}\rangle - D.
\label{S_12n_exact}
\end{equation}
The dissipation term is present in this case, and the pressure
term shares the $same$ derivative factor ${\cal P}_{yv}$ found in
Eq.~(\ref{S_02n+2_exact}). Keeping the first two terms in the mean
field expansion (see the first equation in Sec.~VB), we have for
that term
\begin{equation}
{\cal P}_{yv} = -H{\Delta u_r \Delta v_r \over r} - B{\Delta v_r
\over (P r)^{2\over3}} \label{pyv_expansion}
\end{equation}
where $H$ and $B$ are unknown constants, and $P$, as before, is
the rate of forcing. The second term is chosen in order that the
pressure term in the third-order equation $\overline{{\cal P}_{yv}
\Delta v_r} = 0$, which is a K41 constraint. The two constants $H$
and $B$ are then related through
\begin{equation}
-H S_{1,2} = B S_{0,2} {r^{1/3} \over P^{2\over3}}.
\label{hb_constraint}
\end{equation}
If we insert the model for the pressure into
Eq.~(\ref{S_02n+2_exact}) for $n=1$ (say), we can extract the
value of the constant $H$ from experimental data on the left and
right sides of that equation. With this value of $H$ in
Eq.~(\ref{S_12n_exact}) we have a complete expression for the
pressure term. The remaining imbalance, if any, must be the
dissipation term.

To evaluate the magnitude of the dissipation terms, let us return
to Eq.~(\ref{S_02n+2_exact}), for which, with
\begin{equation}
{\cal P}_{yv} = -H {\Delta u_r\Delta v_r \over r} - B{\Delta v_r
\over (Pr)^{2\over3}}
\end{equation}
we have the following result:
\begin{eqnarray}
-2n\langle {\cal P}_{yv}\Delta u_r(&\Delta v_r&)^{2n-1}\rangle
= -2n(-H {S_{2,2n}\over r} - B {S_{1,2n}\over(Pr)^{2\over3}}) \nonumber\\
&=& 2nH({S_{2,2n}\over r} - {1\over r}{S_{1,2}S_{1,2n}\over
S_{0,2}}). \label{pr_even}
\end{eqnarray}
The first equality in Eq.~(\ref{pr_even}) uses
Eq.~(\ref{pyv_expansion}) and the second equality uses
Eq.~(\ref{hb_constraint}). We extract the value of $H$ by
substituting Eq.~(\ref{pr_even}) in Eq.~(\ref{S_02n+2_exact}) and
solving for it from experimental data for the case $n=1$. Figure
\ref{h} shows $H$ computed in this manner. The value ranges
between $0.3$ and $0.1$ in the inertial range with significant
scatter for larger scales. The value of $H$ can also be computed
from the sixth-order equation (i.e., $n=2$ in
Eq.~(\ref{S_02n+2_exact})) in this same way. Though the statistics
are not as nicely convergent, the result is that $H \approx 0.28$
with deviations of the order of $0.1$. As a further check, we can
compute $H$ from the above equation assuming the scaling exponents
$\xi_{2,2} = \xi_{0,4} = 1.26$ and taking from the data the ratio
$S_{0,4}/S_{2,2} \approx 4.5$ in the inertial range. We obtain $H
\approx 0.37$. Thus, the precise value of $H$ appears to depend on
the order of the moment and on the scale range but, since the
pressure term is relatively small in the inertial range, the exact
choice of $H$ may not be critical in determining the dominant
dissipation term. In any case, the uncertainty in these estimates
does not allow us to be too definitive, and so we shall proceed
with an average value of 0.25.

Our proposal is to substitute this value of $H$ in
Eq.~(\ref{S_12n_exact}) for $n=2$. In detail, that equation is
\begin{eqnarray}
-2n\langle {\cal P}_{yv}(\Delta v_r)^{2n-1}\rangle
&=& -2n(-H {S_{1,2n}\over r} - B {S_{0,2n}\over(Pr)^{2\over3}}) \nonumber\\
&=& 2nH({S_{1,2n}\over r} - {1\over r}{S_{1,2}S_{0,2n}\over
S_{0,2}}). \label{pr_odd}
\end{eqnarray}
This pressure term is seen to account for only about 10\% of the
imbalance of Eq.~(\ref{S_12n_exact}), as shown in Fig.
\ref{press_S1,4}. We now substitute the pressure contribution of
Eq.~(\ref{pr_odd}) back into Eq.~(\ref{S_12n_exact}) in order to
estimate the only unknown term $D$. Figure \ref{diss_S1,4} shows
that the dissipation term so extracted dominates RHS, and is much
larger than the pressure term. The dissipation term alone is
comparable to the entire LHS of the equation.

There is another equation involving fifth-order structure
functions that contains the same form for the pressure gradient as
Eqs.~(\ref{S_02n+2_exact}) and (\ref{S_12n_exact}). It is
Eq.~(\ref{S_32_approx}) including the pressure and dissipation
terms which, in full, reads as
\begin{equation}
{\partial S_{3,2}\over\partial r} + {4\over r}S_{3,2}={8\over
3r}S_{1,4} - 2\langle {\cal P}_{y \nu }(\Delta u_r)^2 \Delta v_r
\rangle + D. \label{S_32_press_incl}
\end{equation}
We now follow a similar procedure as before. For
Eq.~(\ref{S_32_press_incl}), the pressure term according to the
mean field model is
\begin{eqnarray}
-2\langle {\cal P}_{y \nu }(\Delta u_r)^2 \Delta v_r \rangle  &=&
-2(-H{S_{3,2}\over r} - B {S_{2,2}\over (Pr)^{2/3}}) \nonumber\\
&=&2H({S_{3,2}\over r} - {1\over r}{S_{1,2}S_{2,2}\over S_{0,2}}).
\label{pr_s32}
\end{eqnarray}

Figure \ref{rhs_S3,2} shows that the RHS of
Eq.~(\ref{S_32_approx}) balances the LHS up to about 80$\%$ in the
inertial range. The imbalance is due to a possible mix of pressure
and dissipation. The pressure term computed from
Eq.~(\ref{pr_s32}) is shown in Fig.~\ref{press_S3,2}. It makes a
10$\%$ contribution in the inertial range. The dissipation term,
being the remainder ($D = LHS - RHS - I_p$), is plotted in
Fig.~\ref{diss_S3,2}; while it shows significant scatter, it is
clearly small in the inertial range (of the order of 15\% or less)
while increasing, as it must, toward the dissipative scales.

From the above two examples it appears that the overall order of
the structure function (in this case the fifth) is not enough to
prescribe the relative importance of the pressure and dissipation
terms. The equations that relate different $components$ of the
fifth-order structure function tensor: Eq.~(\ref{S_32_press_incl})
shows different ratios of pressure and dissipation terms from
Eq.~(\ref{S_12n_exact}), whereas the overall order of the
structure function in both cases is 5. The former equation seems
to balance more or less without pressure and dissipation while for
the latter equation the pressure term and, particularly, the
dissipation term prove to be essential.

There is one further detail that needs to be mentioned for
completeness. This concerns the relative magnitudes of the $H$ and
$B$ terms in Eq.~(\ref{pr_s32}). The ratio of the $B$ term to $H$
term in Eq.~(\ref{pr_s32}) is about 10$\%$, while this ratio is of
the order of 1$\%$ for Eq.~(\ref{S_02n+2_exact}) and approximately
80$\%$ for Eq.~(\ref{S_12n_exact}). The conclusion seems to be
that the relative importance of the $B$-term depends on the order
of the structure function as well as the component of the
structure function being considered.

\section{Concluding remarks}
Our experimental results are assessed in the context of a mean
field model due to Yakhot. The model allows us to write the
pressure terms which we cannot measure directly, in terms of the
velocity structure functions that we can measure. (The pressure
terms appear here in a different form from those used in
turbulence modeling, and so the value of the present work to that
endeavor is unclear.) Among the assumptions made, the most drastic
one is the use of the same pressure model for $2d$ and $3d$
turbulence.

Nevertheless, if we adopt the pressure model in
Eq.~(\ref{S_02n+2_exact}) in which the dissipation terms are
thought to be negligible (see \cite{yakhot1} and \cite{hill1} for
symmetry and asymptotic arguments as to why this might be so), the
coefficients $H$ and $B$ can be obtained, and thus the pressure
terms can be modeled. We can now proceed to analyze odd-order
equations that have the same structure for pressure terms. Since
the pressure term is known, we can deduce the only remaining term,
namely the dissipation. For one equation (\ref{S_12n_exact}), the
dissipation term is of the order of $80\%$ of the balance. Another
dynamical equation (\ref{S_32_press_incl}) for the same order of
the structure function has a different structure, and there, the
dissipation term is relatively small. This is a new and
interesting statement about the inertial range dynamics, but its
validity depends on the pressure model used. At least one outcome
of the calculations is tautologically correct: in all the cases
considered here, the dissipation range is always dominated by the
dissipation term $D$.

Yakhot's theory postulates the existence of a critical dimension,
$d_c$. This, in itself, is not implausible \cite{frisch}. However,
the analytic structure of the NS equations in the neighborhood of
$d_c$ and the extent of the neighborhood remain unclear. The
theory yields certain exponents for the vicinity of $d_c$, but the
details on which they are based need closer scrutiny; at least to
us, some of the steps remain unclear. Thus, while the numerical
values of the exponents, as well as that of $d_c$ itself, are
probably not to be taken literally, we should be interested in
drawing some qualitative conclusions.

Such conclusions come from a few independent sources. First, the
prediction of the theory for the PDF of $\Delta v_r$ for $2d$
turbulence is in good agreement with simulations and experiments
\cite{boffetta,tabeling}. Second, the conditional expectation of
the pressure terms in $3d$ simulations \cite{gotoh} appear to
follow the mean field theory, at least for modest values of the
velocity increments. Third, shell model calculations \cite{jensen}
show that the behavior expected near the critical dimension can be
observed as one varies a coupling parameter. Finally, the present
comparisons with experimental data at high Reynolds numbers reveal
that the scaling of the PDF of $\Delta v_r$ for small and large
$\Delta v_r$ are in some measure of agreement with the theory. All
these are positive developments. However, since many details are
unclear, it remains to be seen as to whether the theory will
evolve into a rational framework. For now, we find it to be
remarkably interesting and worthy of some attention.

We thank V. Yakhot for numerous helpful discussions without which
this paper would not have been written. However, he should not be
held responsible for our interpretations of the theory. We also
thank R.J. Hill and J. Schumacher for helpful comments, and the
latter also for reading the manuscript carefully. The work was
supported by the US Office of Naval Research.

\begin{table}
\begin{tabular}{|c|c|c|c|}
Order &  DNS   &  Experiment  &  From Eq.~(\ref{y_exp}) \\ \hline
-0.80 & -0.317 & -0.313 &-0.328 \\\hline -0.20& -0.077 & -0.078
&-0.079\\\hline 0.10 & 0.036 & 0.039 &0.039\\\hline 0.20 & 0.073 &
0.076 &0.077\\\hline 0.30 & 0.112 & 0.113 &0.115\\\hline 0.40 &
0.150 & 0.150 &0.153\\\hline 0.50 & 0.187 & 0.190 &0.190\\\hline
0.60 & 0.223 & 0.221 &0.227\\\hline 0.70 & 0.260 & 0.265
&0.263\\\hline 0.80 & 0.296 & 0.292 &0.299\\\hline 0.90 & 0.332 &
0.333 &0.335\\\hline 1.00 & 0.366 & 0.372 &0.370\\\hline 1.25 &
0.452 & 0.458 &0.456\\\hline 1.50 & 0.536 & 0.542 &0.540\\\hline
1.75 & 0.619 & 0.628 &0.622\\\hline 2 & 0.699 & 0.708
&0.701\\\hline 3 & 1 & 1 &1\\\hline 4 & 1.279 & 1.26
&1.271\\\hline 5 & 1.536 & 1.56 & 1.517\\\hline 6 & 1.772 & 1.71
&1.742\\\hline 7 & 1.989 & 1.97 &1.948\\\hline 8 & 2.188 & 2.05
&2.138\\\hline 9 & 2.320 & 2.20 &2.314\\\hline 10& 2.451 & 2.38 &
2.477
\end{tabular}
\caption{Comparison of exponents from the DNS data and the
experiment (both using ESS) and Yakhot's formula,
Eq.~(\ref{y_exp}).}
\end{table}
\begin{figure}
\epsfxsize=8.5truecm \epsfysize=8truecm \epsfbox{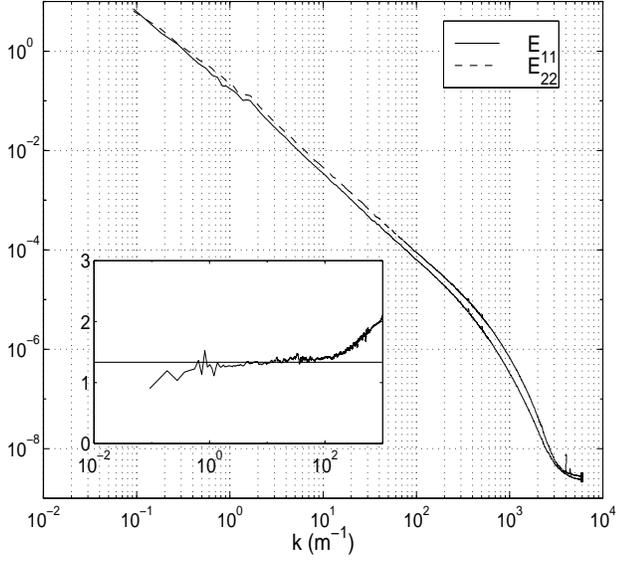}
\caption{Longitudinal and transverse spectral densities, Taylor
microscale Reynolds number $\approx 19,000$.} \label{bd_spect}
\end{figure}
\begin{figure}
\epsfxsize=8.5truecm \epsfysize=8truecm \epsfbox{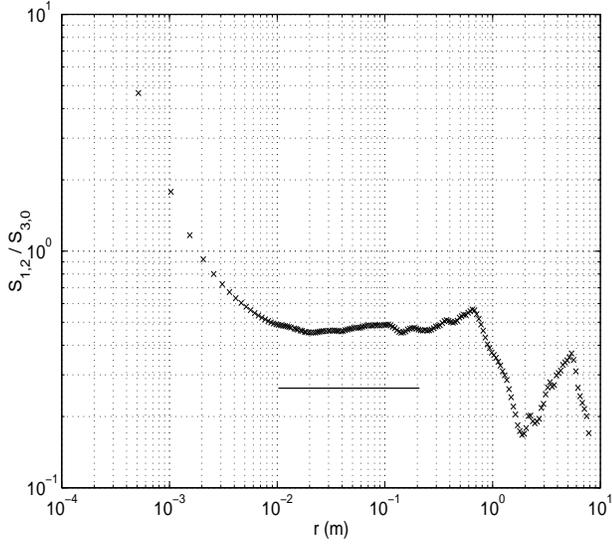}
\caption{The ratio of $S_{1,2}$ to $S_{3,0}$ showing some
deviations from the isotropic value of $1/3$ in the inertial
range. In this figure and elsewhere, the horizontal bar in the
figure represents a conservative extent of the scaling region
determined from Kolmogorov's 4/5-ths law.} \label{s12bys3}
\end{figure}
\begin{figure}
\epsfxsize=8.5truecm \epsfysize=8truecm \epsfbox{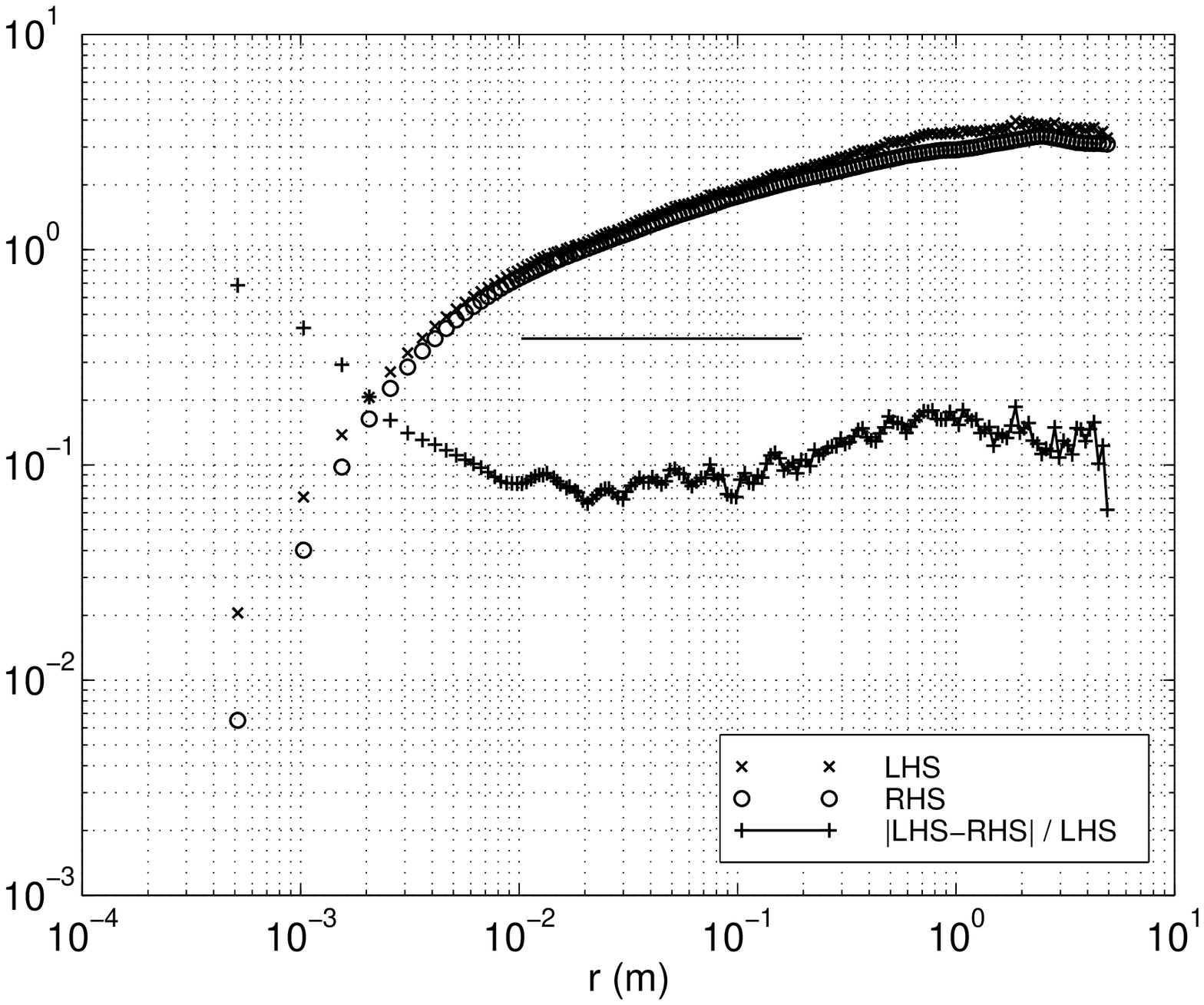}
\caption{The terms of Eq.~(\ref{S_40_approx}) with their relative
difference.} \label{s40}
\end{figure}
\begin{figure}
\epsfxsize=8.5truecm \epsfysize=8truecm \epsfbox{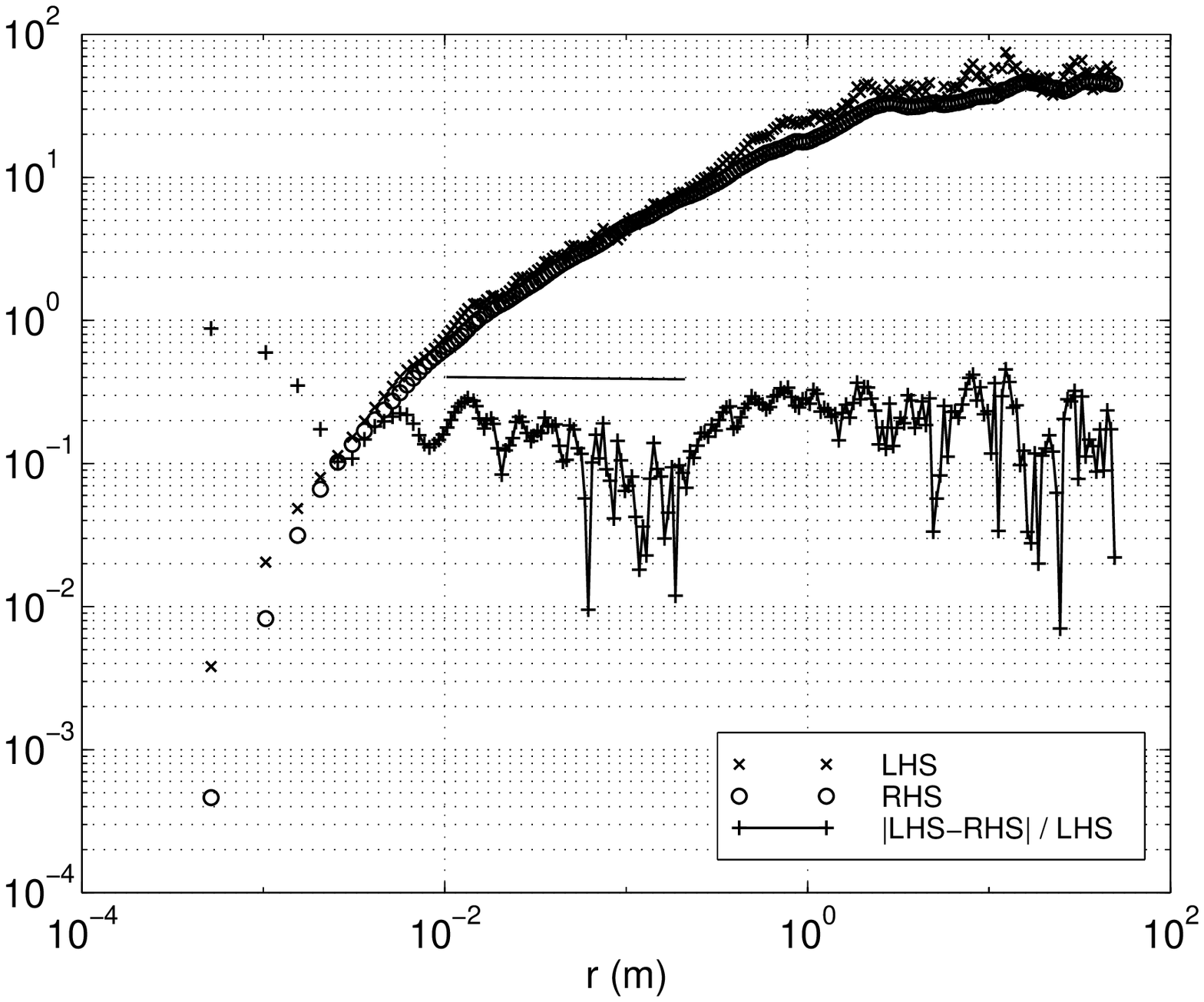}
\caption{The terms of Eq.~(\ref{S_60_approx}) with their relative
difference.} \label{s60}
\end{figure}
\begin{figure}
\epsfxsize=8.5truecm \epsfysize=8truecm \epsfbox{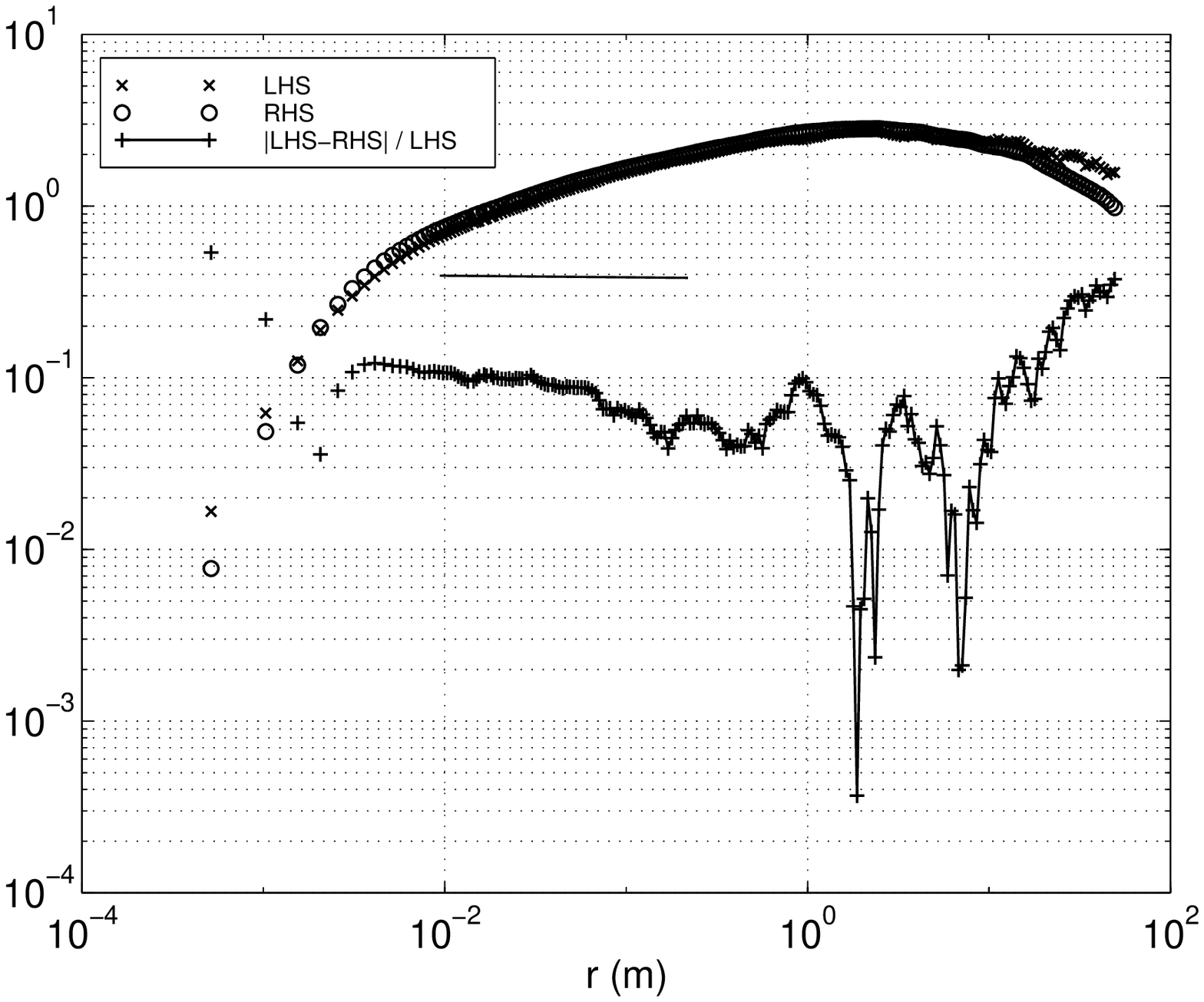}
\caption{The terms of Eq.~(\ref{S_04_approx}) with their relative
difference.} \label{s04}
\end{figure}
\begin{figure}
\epsfxsize=8.5truecm \epsfysize=8truecm \epsfbox{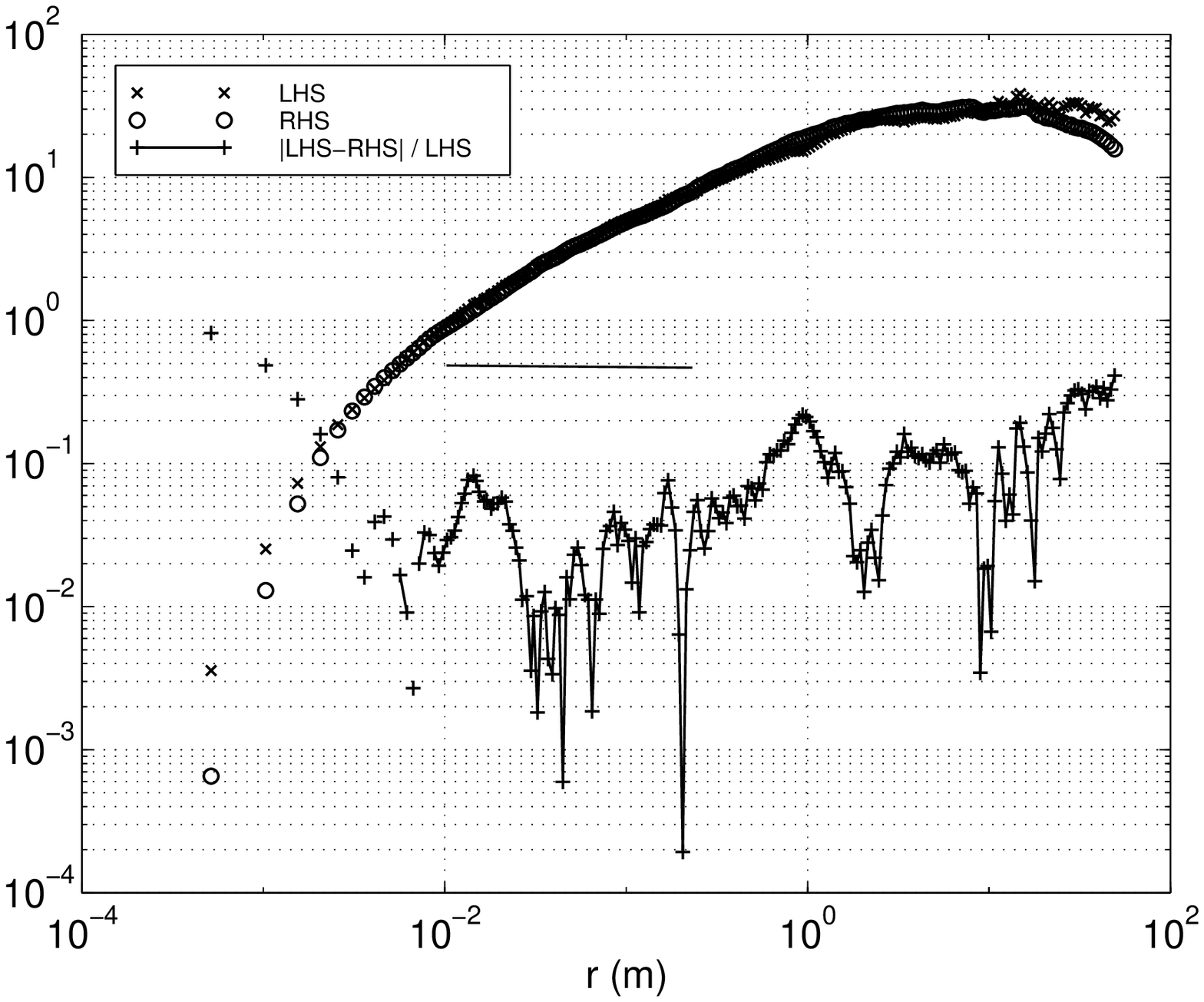}
\caption{The terms of Eq.~(\ref{S_06_approx}) with their relative
difference.} \label{s06}
\end{figure}
\begin{figure}
\epsfxsize=8.5truecm \epsfysize=8truecm \epsfbox{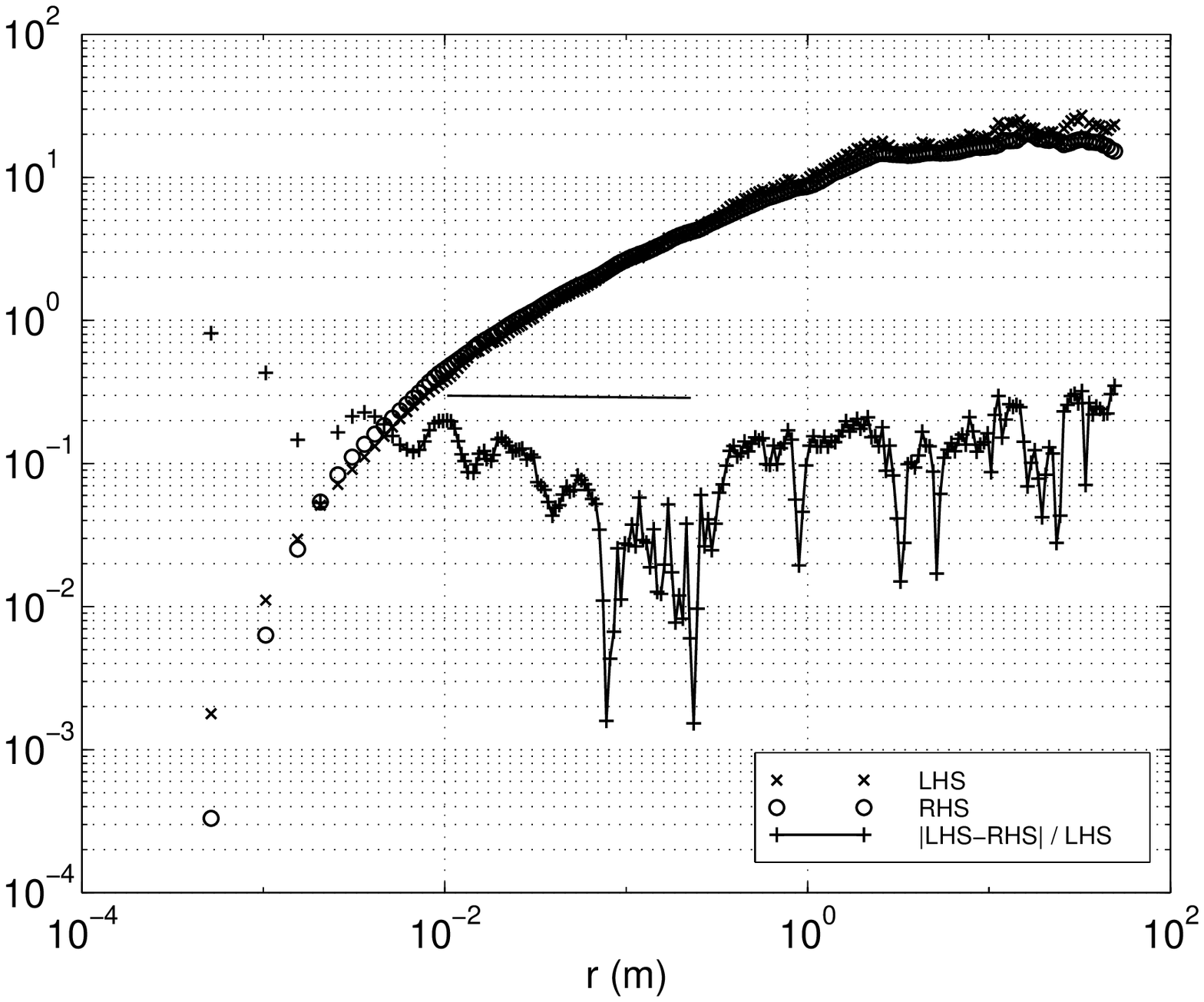}
\caption{The terms of Eq.~(\ref{S_mixed6_approx}) with their
relative difference.} \label{sm6}
\end{figure}
\begin{figure}
\epsfxsize=8.5truecm \epsfysize=8truecm \epsfbox{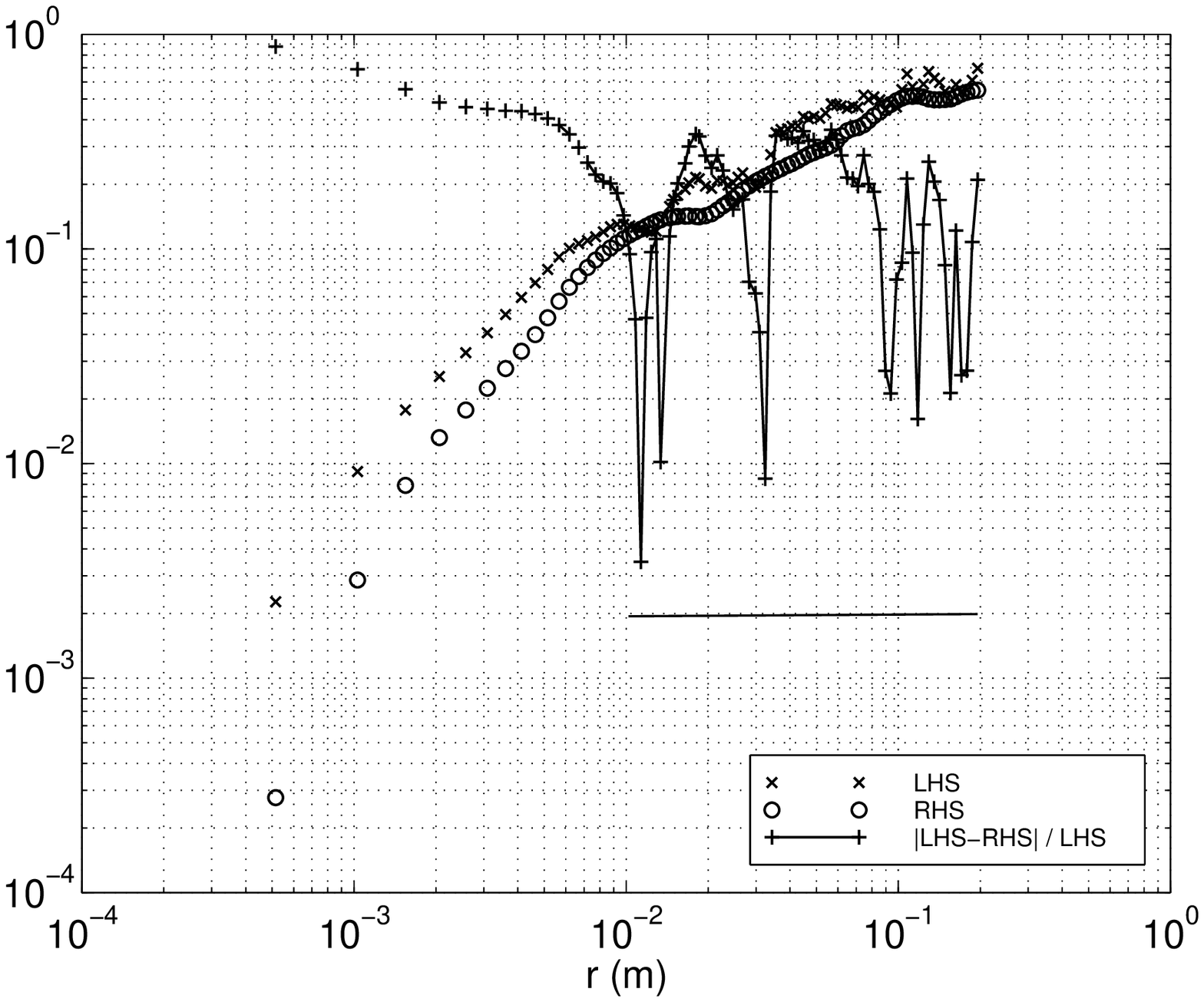}
\caption{The terms of Eq.~(\ref{S_50_approx}) with their relative
difference.} \label{s50}
\end{figure}
\begin{figure}
\epsfxsize=8.5truecm \epsfysize=8truecm \epsfbox{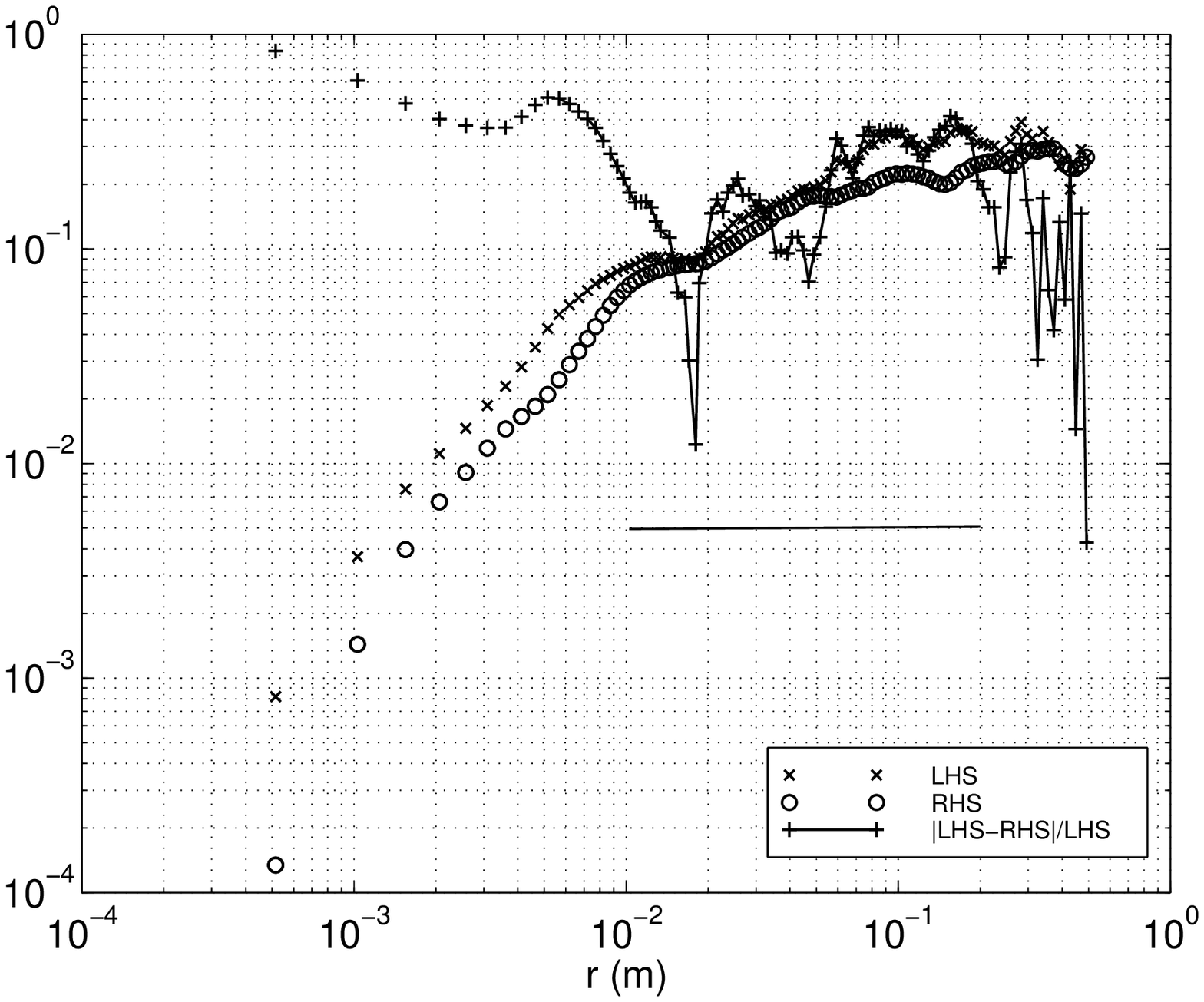}
\caption{The terms of Eq.~(\ref{S_32_approx}) with their relative
difference.} \label{s32}
\end{figure}
\begin{figure}
\epsfxsize=8.5truecm \epsfysize=8truecm \epsfbox{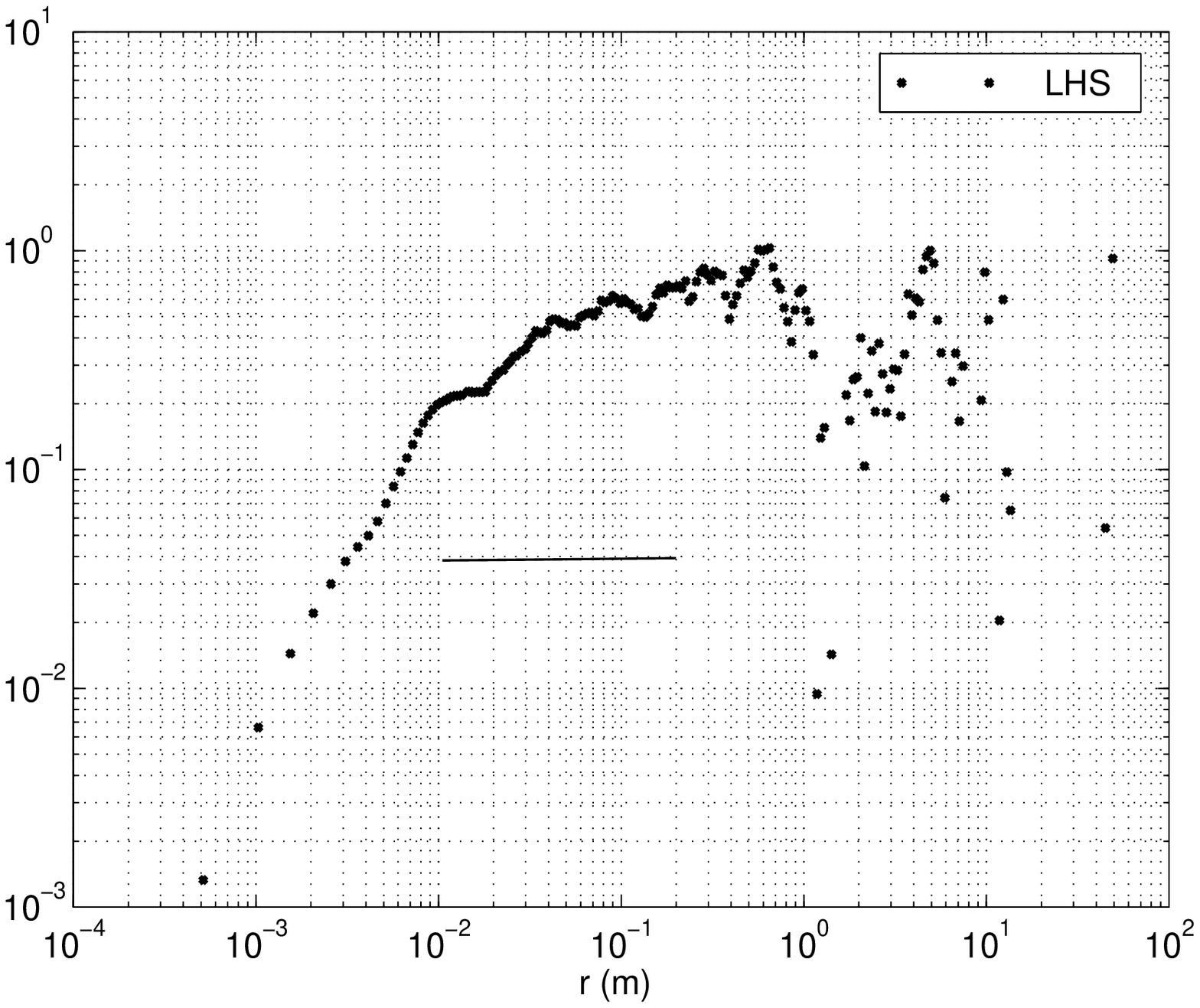}
\caption{The terms of Eq.~(\ref{S_14_approx}) with their relative
difference.} \label{s14}
\end{figure}
\begin{figure}
\epsfxsize=8.5truecm \epsfysize=8truecm
\epsfbox{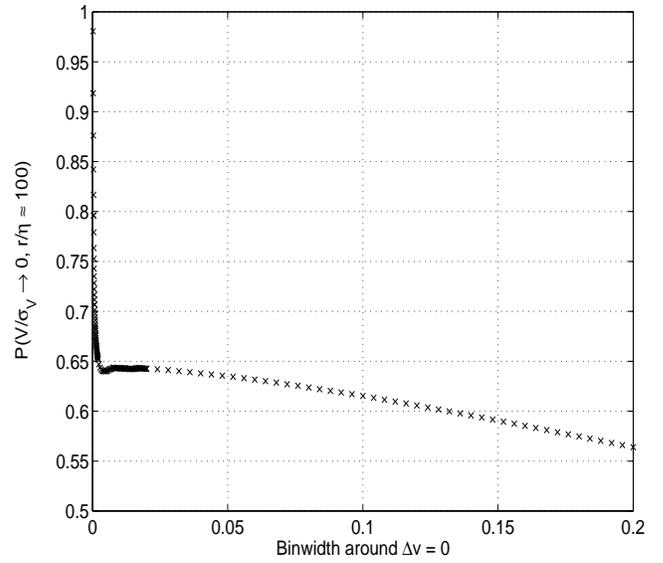} 
\caption{The value of the PDF
as $V/\sigma_V \rightarrow 0$ for $r \approx 100\eta$ for
different binwidths.} \label{find_peak}
\end{figure}
\begin{figure}
\epsfxsize=8.5truecm \epsfysize=8truecm \epsfbox{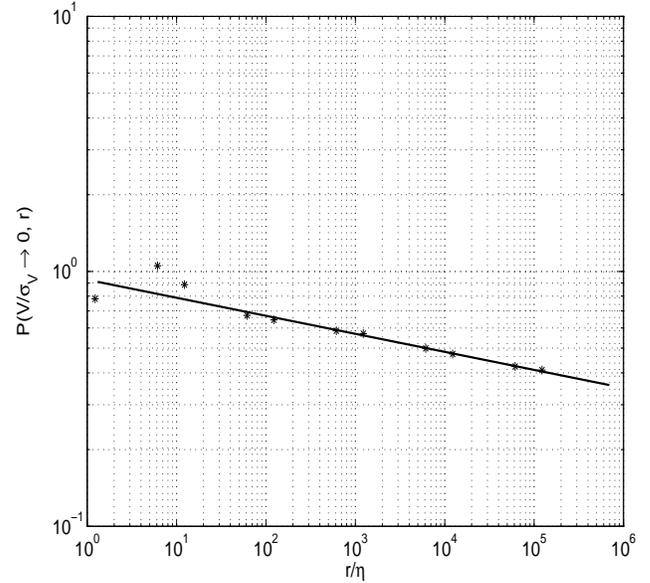}
\caption{Log-log plot of the peak values of the PDFs of transverse
velocity increments. The line indicates a slope of -0.065.}
\label{peakpdfs}
\end{figure}
\begin{figure}
\epsfxsize=8.5truecm \epsfysize=8truecm
\epsfbox{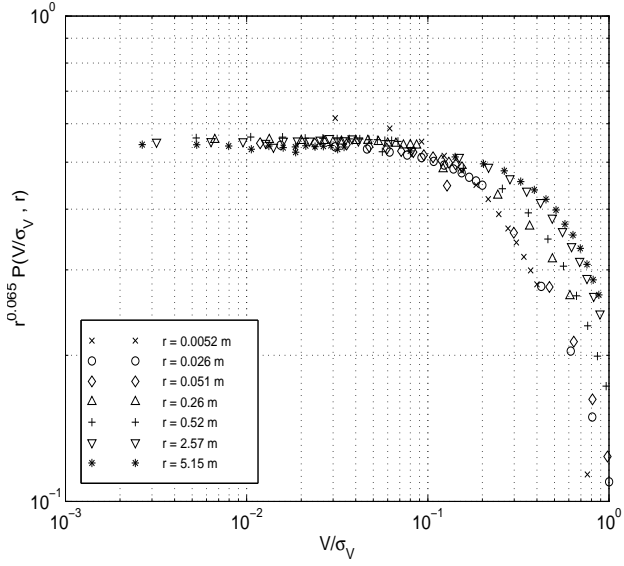} \caption{Log-log plot of the near-peak
values of the PDFs of transverse velocity increments. The collapse
of the data occurs for the normalization $r^\kappa P(V/r^\kappa)$
where $\kappa=0.065$.} \label{nearpeak}
\end{figure}
\begin{figure}
\epsfxsize=8.5truecm \epsfysize=8truecm \epsfbox{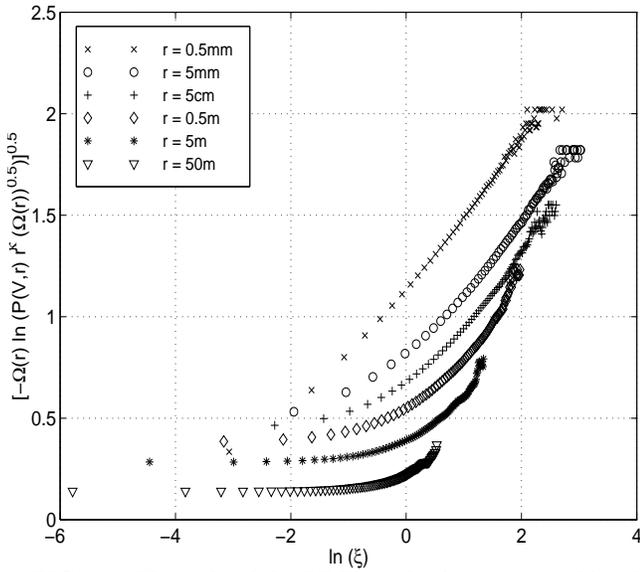}
\caption{The tails of the PDFs in the form required by the theory
(see Eq.~(\ref{pdf_anl2}). Lognormality of the tails requires that
the data follow straight lines.} \label{pdf_ncoll}
\end{figure}
\begin{figure}
\epsfxsize=8.5truecm \epsfysize=8truecm \epsfbox{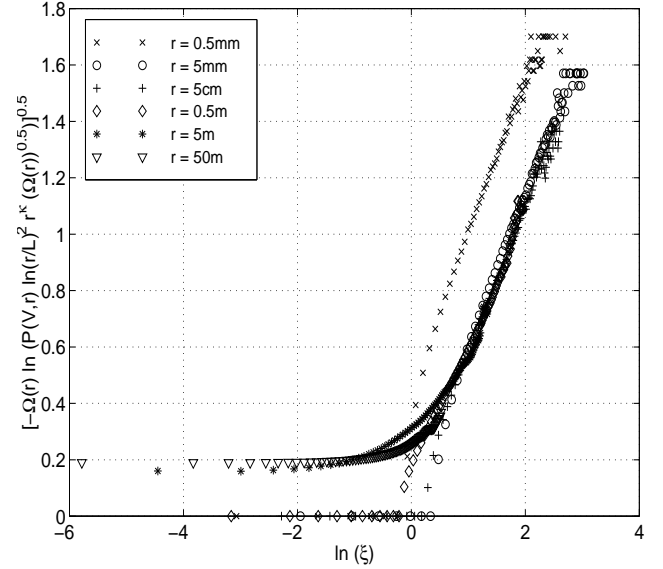}
\caption{The collapsed tails of the PDF from the previous plot,
taking into account an {\it ad hoc} prefactor of $(\ln(r/L))^2$.}
\label{pdf_coll}
\end{figure}
\begin{figure}
\epsfxsize=8.5truecm \epsfysize=8truecm \epsfbox{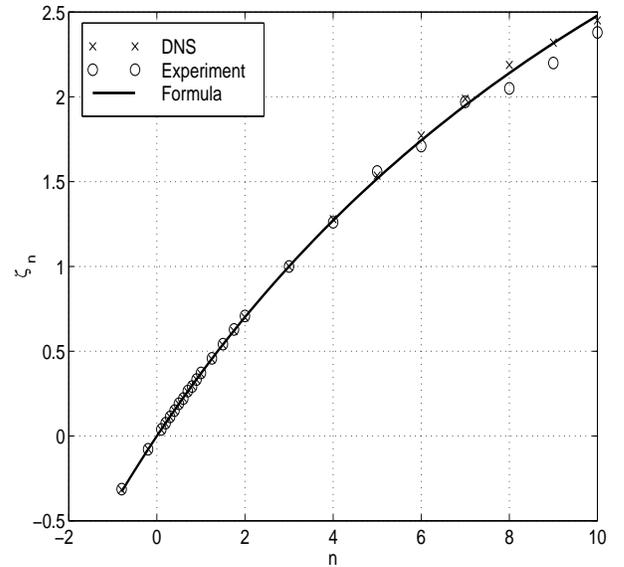}
\caption{Comparison of the DNS and experimental values of
$\zeta_n$ with those from Eq.~(\ref{y_exp}). The numbers in the
plot are for longitudinal structure functions, but these are
identical to those for transverse structure functions in the
isotropic sector see \protect\cite{arad,kurien}.} \label{exps}
\end{figure}
\begin{figure}
\epsfxsize=8.5truecm \epsfysize=8truecm \epsfbox{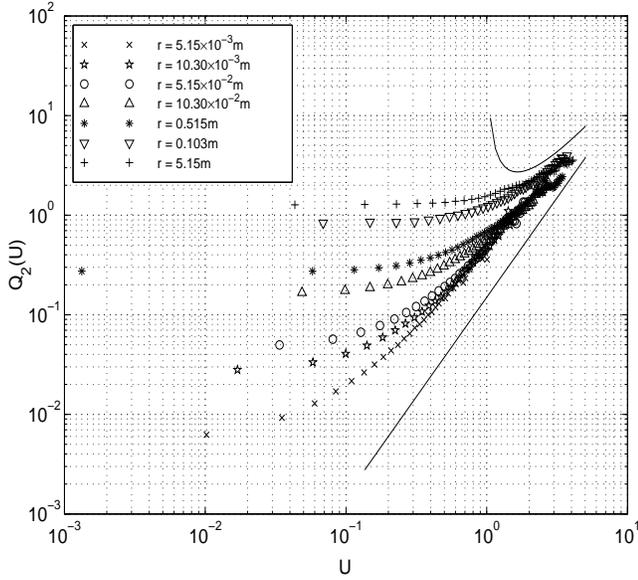}
\caption{$Q_2(U)$, the conditional expectation value of $V^2$ on
$U$ for various different $r$, on a log-log scale. The upper solid
line indicates $U^2/log(U^2)$, the lower solid line indicates the
$U^2$ scaling slope.} \label{q2plot}
\end{figure}
\begin{figure}
\epsfxsize=8.5truecm \epsfysize=8truecm \epsfbox{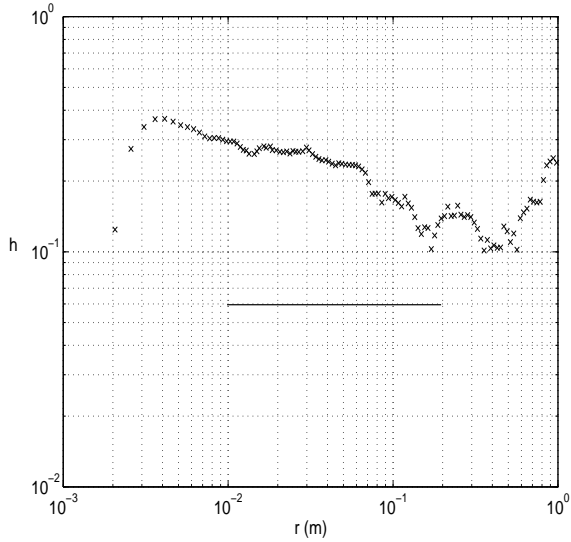}
\caption{The coefficient $H$ of the pressure term computed by
substituting Eq.~(\ref{pr_even}) in Eq.~(\ref{S_02n+2_exact}) with
$n=1$.} \label{h}
\end{figure}
\begin{figure}
\epsfxsize=8.5truecm \epsfysize=8truecm
\epsfbox{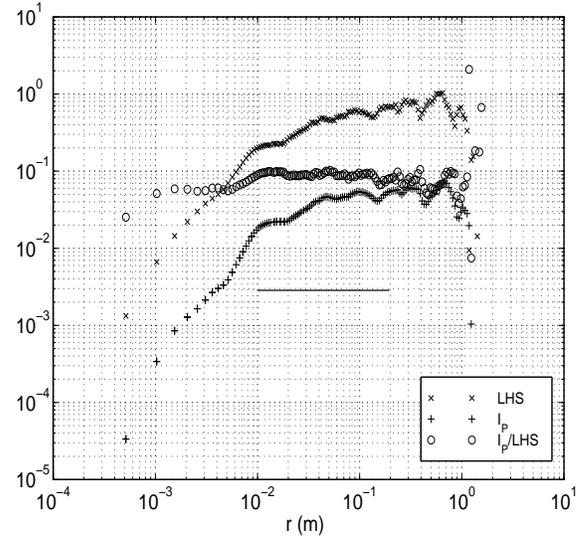} \caption{The LHS and pressure terms
of Eq.~(\ref{S_12n_exact}) with $n=2$ and $H=0.25$ in the pressure
model; the ratio indicates that the pressure term as computed from
the model is about 10$\%$ of the balance.} \label{press_S1,4}
\end{figure}
\begin{figure}
\epsfxsize=8.5truecm \epsfysize=8truecm \epsfbox{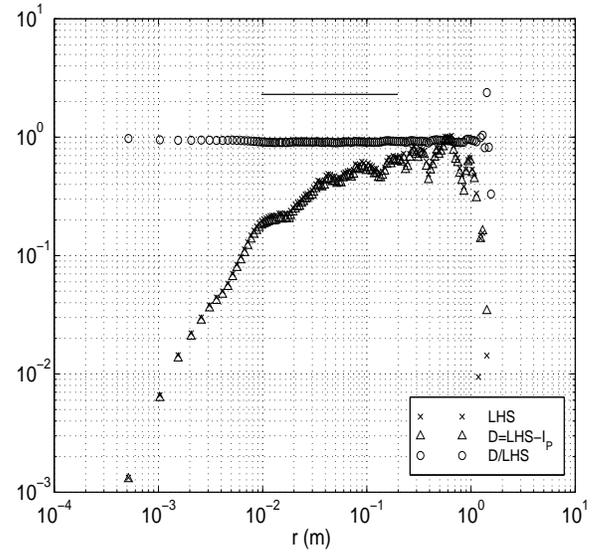}
\caption{The terms of Eq.~\ref{S_12n_exact} with $n=2$, $H=0.25$;
the difference LHS-RHS indicates the magnitude of the dissipation
term.} \label{diss_S1,4}
\end{figure}
\begin{figure}
\epsfxsize=8.5truecm \epsfysize=8truecm \epsfbox{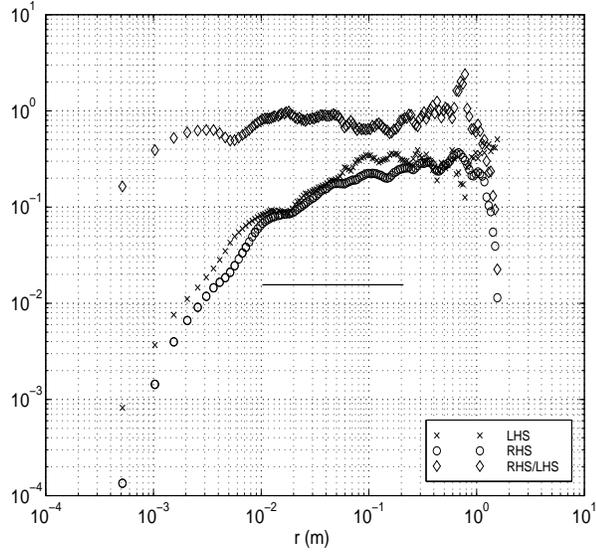}
\caption{The LHS and RHS of the approximate equation,
Eq.~(\ref{S_32_approx}), without pressure or dissipation terms.
The RHS balances the equation up to about 80$\%$ in the inertial
range, less as we approach the dissipation and large scales.}
\label{rhs_S3,2}
\end{figure}
\begin{figure}
\epsfxsize=8.5truecm \epsfysize=8truecm
\epsfbox{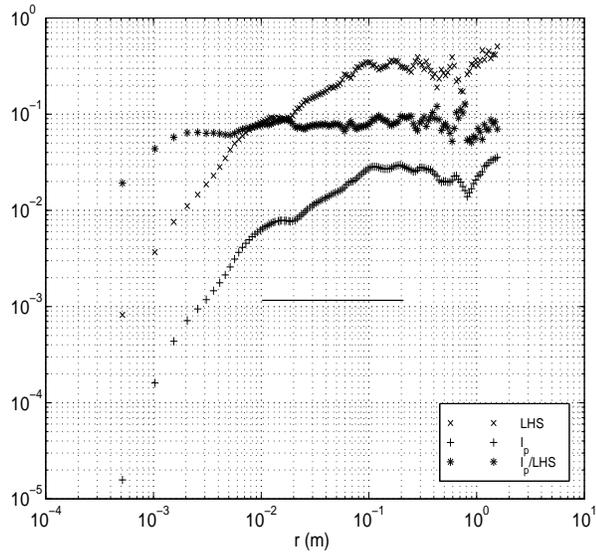} \caption{The LHS and pressure terms
$I_p$ of Eq.~(\ref{S_32_press_incl}) with their ratio showing that
the pressure only accounts for about 8-10$\%$ of the balance, less
as one approaches the dissipative scales and large scales.}
\label{press_S3,2}
\end{figure}
\begin{figure}
\epsfxsize=8.5truecm \epsfysize=8truecm \epsfbox{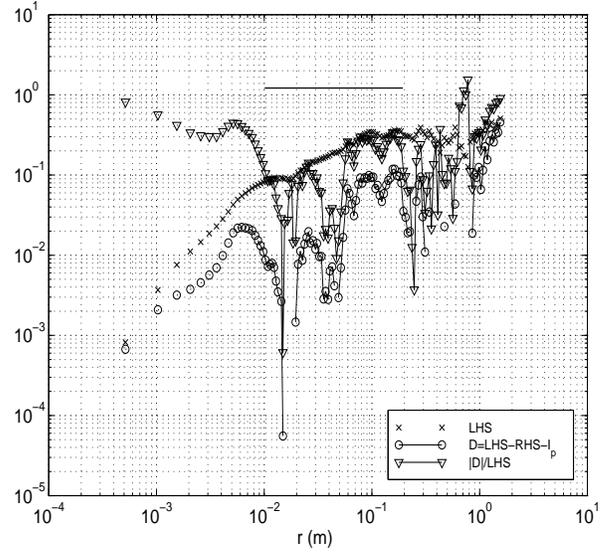}
\caption{The LHS and dissipation terms $D$ of
Eq.~(\ref{S_32_press_incl}) with the dissipation computed by
subtracting the RHS and the pressure (see Fig.~\ref{press_S3,2})
from LHS; the ratio of $D$ to LHS indicates a large scatter but
shows the dissipation contributes in the inertial range only to
about 10$\%$, but that it significantly increases towards the
dissipative scales.} \label{diss_S3,2}
\end{figure}
\end{document}